\documentclass{SciPost}

\hypersetup{
    colorlinks,
    linkcolor={red!50!black},
    citecolor={blue!50!black},
    urlcolor={blue!80!black}
}

\usepackage[bitstream-charter]{mathdesign}
\DeclareSymbolFont{usualmathcal}{OMS}{cmsy}{m}{n}
\DeclareSymbolFontAlphabet{\mathcal}{usualmathcal}

\fancypagestyle{SPstyle}{
\fancyhf{}
\lhead{\colorbox{scipostblue}{\bf \color{white} ~SciPost Physics }}
\rhead{{\bf \color{scipostdeepblue} ~Submission }}

\fancyfoot[C]{\textbf{\thepage}}
}

\usepackage{changepage} 

\begin{document}

\pagestyle{SPstyle}

\newcommand{\atopgen}[2]{\genfrac{}{}{0pt}{1}{#1}{#2}}

\newcommand{\TMT}{TMM}
\newcommand{\ETMT}{ETMA}
\newcommand{\RI}{RI}
\newcommand{\RP}{RP}
\newcommand{\TR}{TR}
\newcommand{\CW}{CW}
\newcommand{\EW}{EW}
\newcommand{\GEDF}{GEDF}
\newcommand{\GF}{GF}
\newcommand{\CC}{CC}
\newcommand{\IBF}{IBF}
\newcommand{\VLBF}{VLBF}
\newcommand{\CE}{CE}
\newcommand{\EpsRel}{\bar{\varepsilon}_{\mbox{\scriptsize r}}}
\newcommand{\MuRelj}[1]{\mu_{\mbox{\scriptsize r}#1}}
\newcommand{\Esj}[1]{E_{\mbox{\scriptsize s}#1}}
\newcommand{\Bpj}[1]{B_{\mbox{\scriptsize p}#1}}
\newcommand{\zrj}[1]{z_{\mbox{\scriptsize r}#1}}
\newcommand{\Esr}{E_{\mbox{\scriptsize sr}}}
\newcommand{\Bpr}{B_{\mbox{\scriptsize pr}}}
\newcommand{\EsjPM}[2]{E_{\mbox{\scriptsize s}#1}^{#2}}
\newcommand{\BpjPM}[2]{B_{\mbox{\scriptsize p}#1}^{#2}}
\newcommand{\EpjPM}[2]{E_{\mbox{\scriptsize p}#1}^{#2}}
\newcommand{\EsjPls}[1]{\EsjPM{#1}{+}}
\newcommand{\EsjMin}[1]{\EsjPM{#1}{-}}
\newcommand{\BpjPls}[1]{\BpjPM{#1}{+}}
\newcommand{\BpjMin}[1]{\BpjPM{#1}{-}}
\newcommand{\EpjPls}[1]{\EpjPM{#1}{+}}
\newcommand{\LEWC}[1]{l_{\mbox{\scriptsize EWc}}^{(#1)}}
\newcommand{\Labs}[1]{l_{\mbox{\scriptsize abs}}^{(#1)}}
\newcommand{\Lcoh}{l_{\mbox{\scriptsize coh}}}
\newcommand{\TILS}{TILS}
\newcommand{\PhiBulk}{\phi_{\mbox{\scriptsize eq}}}
\newcommand{\PhiPeak}{\varphi_0^{\mbox{\scriptsize (peak)}}}
\newcommand{\PhiSPR}{\varphi_0^{\mbox{\scriptsize (SPR)}}}
\newcommand{\RIBulk}{n^{\mbox{\scriptsize (bulk)}}}
\newcommand{\RIContact}{n^{\mbox{\scriptsize (cnt)}}}
\newcommand{\RISPol}{n_{\mbox{\scriptsize s}}}
\newcommand{\RIPPol}{n_{\mbox{\scriptsize p}}}
\newcommand{\RIWater}{n_{\mbox{\scriptsize water}}}
\newcommand{\RIAir}{n_{\mbox{\scriptsize air}}}
\newcommand{\OfOrder}{\mathcal{O}}

\newcommand{\PhiCBulk}{\phiTRjj{\mbox{\scriptsize bulk}}}
\newcommand{\PhiCntS}{\phiTRjj{\mbox{\scriptsize cnt,s}}}
\newcommand{\PhiCntP}{\phiTRjj{\mbox{\scriptsize cnt,p}}}
\newcommand{\CritAngNemS}{\varphi_{0c}^{\mbox{\scriptsize (nem,s)}}}
\newcommand{\CritAngNemP}{\varphi_{0c}^{\mbox{\scriptsize (nem,p)}}}
\newcommand{\CritAngIso}{\varphi_{0c}^{\mbox{\scriptsize (iso)}}}
\newcommand{\IScat}{I_{\mbox{\scriptsize scat}}}
\newcommand{\LCW}{l_{\mbox{\scriptsize cw}}}
\newcommand{\lambdaCW}{\lambda_{\mbox{\scriptsize cw}}}
\newcommand{\omegaCW}{\omega_{\mbox{\scriptsize cw}}}

\newcommand{\PhiRed}{\phi''_{\mbox{\scriptsize red}}}
\newcommand{\PhiRedC}{\phi''_{\mbox{\scriptsize red,0}}}

\newcommand{\EsrjPM}[2]{E_{\mbox{\scriptsize s}#1}^{#2}}
\newcommand{\BprjPM}[2]{B_{\mbox{\scriptsize p}#1}^{#2}}
\newcommand{\EsrjPls}[1]{\EsrjPM{#1}{+}}
\newcommand{\EsrjMin}[1]{\EsrjPM{#1}{-}}
\newcommand{\BprjPls}[1]{\BprjPM{#1}{+}}
\newcommand{\BprjMin}[1]{\BprjPM{#1}{-}}

\newcommand{\basej}[2]{b_{#1}^{#2}}
\newcommand{\bjPls}[1]{b_{#1}^{+}}
\newcommand{\bjMin}[1]{b_{#1}^{-}}
\newcommand{\bjType}[3]{^{\mbox{\scriptsize (#3)}}b_{#1}^{#2}}
\newcommand{\bjTet}[2]{\bjType{#1}{#2}{ET}}
\newcommand{\bjTct}[2]{\bjType{#1}{#2}{CT}}
\newcommand{\bjTee}[2]{\bjType{#1}{#2}{EE}}
\newcommand{\bjTce}[2]{\bjType{#1}{#2}{CE}}
\newcommand{\bjTsv}[2]{\bjType{#1}{#2}{SV}}
\newcommand{\bjTtv}[2]{\bjType{#1}{#2}{TV}}

\newcommand{\TraMat}[3]{\underline{T}_{#1,#2}^{\mbox{\scriptsize #3}}}
\newcommand{\TraMatEl}[4]{\left(T_{#1,#2}^{\mbox{\scriptsize #4}}\right)_{#3}}
\newcommand{\TraMatS}[2]{\TraMat{#1}{#2}{S}}
\newcommand{\TraMatP}[2]{\TraMat{#1}{#2}{P}}
\newcommand{\TraMatSEl}[3]{\TraMatEl{#1}{#2}{#3}{S}}
\newcommand{\TraMatPEl}[3]{\TraMatEl{#1}{#2}{#3}{P}}
\newcommand{\Epsj}[1]{\varepsilon_{\mbox{\scriptsize r}#1}}
\newcommand{\Esjp}[1]{E_{\mbox{\scriptsize s}#1}^+}
\newcommand{\Esjm}[1]{E_{\mbox{\scriptsize s}#1}^-}
\newcommand{\Esjpm}[1]{E_{\mbox{\scriptsize s}#1}^\pm}
\newcommand{\Bpjp}[1]{B_{\mbox{\scriptsize p}#1}^+}
\newcommand{\Bpjm}[1]{B_{\mbox{\scriptsize p}#1}^-}
\newcommand{\Bpjpm}[1]{B_{\mbox{\scriptsize p}#1}^\pm}
\newcommand{\kzj}[1]{k_{#1}}
\newcommand{\Mat}[3]{\underline{M}^{\mbox{\scriptsize #1},#2,#3}}
\newcommand{\MatEl}[4]{m_{#4}^{\mbox{\scriptsize #1},#2,#3}}
\newcommand{\phiTR}{\varphi_{0c}}
\newcommand{\phiBj}[1]{\varphi_{0\mbox{\scriptsize B}}^{(#1)}}
\newcommand{\phiTRjj}[1]{\varphi_{0\mbox{\scriptsize c}}^{(#1)}}
\newcommand{\DltPhiBaseS}{{\Delta\varphi}_{\mbox{\scriptsize s}0}^{\mbox{\scriptsize (base)}}}
\newcommand{\DltPhiBaseP}{{\Delta\varphi}_{\mbox{\scriptsize p}0}^{\mbox{\scriptsize (base)}}}
\newcommand{\DltPhiHWHM}{{\Delta\varphi}_{0}^{\mbox{\scriptsize (hm)}}}
\newcommand{\phiTRj}{\phiTRjj{j}}
\newcommand{\gamj}[1]{\gamma_{#1}}
\newcommand{\zej}[1]{z_{#1}}
\newcommand{\zcj}[1]{z_{\mbox{\scriptsize C}#1}}

\renewcommand{\Re}{\mbox{Re}}
\renewcommand{\Im}{\mbox{Im}}

\newcommand{\phiO}{\varphi}
\newcommand{\phiX}{\varphi^{(X)}}
\newcommand{\psiR}{\psi^{(r)}}
\newcommand{\psiT}{\psi^{(t)}}

\newcommand{\bareps}[1]{\bar{\varepsilon}_{#1}}
\newcommand{\Jfree}{\vec{J}_{\mbox{\scriptsize free}}}
\newcommand{\rhofree}{\rho_{\mbox{\scriptsize free}}}

\newcommand{\EpsNull}{\varepsilon_0}
\newcommand{\MuNull}{\mu_0}
\newcommand{\Muriel}{\mu_{\mbox{\scriptsize r}}}
\newcommand{\phiB}{\varphi_{\mbox{\scriptsize B}}}
\newcommand{\BA}{BA}
\newcommand{\AI}{AI}
\newcommand{\MPI}{MPI}
\newcommand{\PT}{PT}
\newcommand{\RT}{RT}
\newcommand{\DWBA}{DWBA}
\newcommand{\UP}{UP}
\newcommand{\EOPT}{EOPT}
\newcommand{\FPT}{FPT}
\newcommand{\WDRGXRCGA}{WDRGXRCGA}
\newcommand{\PCCP}{PCCP}
\newcommand{\BOC}{BOC}
\newcommand{\FAC}{FAC}
\newcommand{\DRY}{DRY}
\newcommand{\WET}{WET}
\newcommand{\DD}{DD}
\newcommand{\MAME}{MAME}
\newcommand{\MIME}{MIME}
\newcommand{\CR}{CR}
\newcommand{\SRID}{SRID}
\newcommand{\ME}{ME}
\newcommand{\GFF}{GFF}
\newcommand{\GISAXS}{GISAXS}
\newcommand{\SAXS}{SAXS}
\newcommand{\WAXS}{WAXS}
\newcommand{\GISANS}{GISANS}
\newcommand{\EWDLS}{EWDLS}
\newcommand{\DLS}{DLS}
\newcommand{\SHG}{SHG}
\newcommand{\SP}{SP}
\newcommand{\SPR}{SPR}
\newcommand{\ATR}{ATR}
\newcommand{\DPhsS}[1]{\Delta_{\mbox{\scriptsize s}#1}}
\newcommand{\DPhsP}[1]{\Delta_{\mbox{\scriptsize p}#1}}

%
%

\begin{center}{\Large \textbf{\color{scipostdeepblue}{
Divergence of Light Wave Amplitudes\\in an Interface Layer at Critical Conditions
}}}\end{center}


\begin{center}
R. E. Sigel\textsuperscript{1$\star$}
\end{center}

\begin{center}
{\bf 1} Independent Scientist, 88677 Markdorf, Germany
* res314159@aol.de
\end{center}



\section*{\color{scipostdeepblue}{Abstract}}
\textbf{\boldmath{%
The amplitudes of light modes in a homogeneous interface layer are investigated around the critical conditions (\CC) in a total reflection geometry. \CC\ occur when the normal wave vector vanishes; the resulting divergence upon variation of the angle of incidence is characterized by a critical exponent $-0.5$. Absorption replaces the divergence with a finite peak whose width and height are derived analytically. The high relevance of the amplification for Evanescent Wave Dynamic Light Scattering (\EWDLS) is demonstrated using published experimental data. 
The relation to surface plasmon resonance (\SPR) is briefly discussed. An outlook connects the amplitude divergence to a critical analysis of the Distorted Wave Born Approximation (\DWBA) presented in a companion paper.
}}

\vspace{\baselineskip}

\noindent\textcolor{white!90!black}{%
\fbox{\parbox{0.975\linewidth}{%
\textcolor{white!40!black}{\begin{tabular}{lr}%
  \begin{minipage}{0.6\textwidth}%
    {\small Copyright attribution to authors. \newline
    This work is a submission to SciPost Physics. \newline
    License information to appear upon publication. \newline
    Publication information to appear upon publication.}
  \end{minipage} & \begin{minipage}{0.4\textwidth}
    {\small Received Date \newline Accepted Date \newline Published Date}%
  \end{minipage}
\end{tabular}}
}}
}


\vspace{10pt}
\noindent\rule{\textwidth}{1pt}
\tableofcontents
\noindent\rule{\textwidth}{1pt}
\vspace{10pt}



\section{Introduction} 
\label{Introduction}

Total reflection (TR) geometries — both internal and external — are widely used to confine illumination to a thin boundary layer of a sample. Under suitable conditions, the evanescent wave (EW) penetrates the sample only over a fraction of the wavelength, making experiments sensitive to the interface region.
Applications range from optical methods like Surface Plasmon Resonance (\SPR) \cite{Raether1988SurfacePlasmons,SurfPlasRes,YESUDASU2021}, second harmonic generation (\SHG) \cite{Shen1989,Fiebig05}, Optical Waveguide Light-mode Spectroscopy (OWLS) \cite{Yu2012,Szekacs2013}, or Evanescent Wave Dynamic Light Scattering (\EWDLS) \cite{EWDLS1986,COCIS}, over the X-ray technique Grazing Incidence Small Angle X-ray Scattering (\GISAXS) \cite{Mahmood2020,Smilgies2022} to experiments with neutrons within Grazing Incidence Small Angle Neutron Scattering (\GISANS) \cite{MuellerBuschbaum2013}.
Soft matter and bio-interfaces are targets of particular interest.

Real interfaces are not sharp transitions between bulk phases. Adsorbed layers form refractive index (\RI) profiles that vary continuously across the interface. The relative permittivity (\RP) profile $\EpsRel(z)$ encodes the van der Waals forces that govern cohesive energy, de-mixing, interfacial tension, and related phenomena. Optical experiments sensitive to this profile therefore constitute local probes of the interaction strengths within the interface layer — a perspective that may lead to a deeper understanding of interfaces as intermediaries between two bulk phases.



For a quantitative evaluation of interface-sensitive optical experiments, a theoretical modeling of light propagation in an interface layer is required. A reliable workhorse to handle the \RI\ profile on a simple and efficient level is a transfer matrix method (\TMT) for the applied radiation \cite{LEKNER}. Technically, the \TMT\ describes the propagation in a layered profile, where conventionally two complex exponentials (\CE s) are employed as basis functions in the representation of the field. With a sufficiently fine layer spacing, a \TMT\ forms a good approximation for a continuous profile.
However, there are critical conditions (\CC) in a \TR\ geometry where conventional \TMT s break down and become singular \cite{TRAMATSINC}. \CC\ are realized when the normal wave vector component $k_j$ in layer $j$ is equal to zero. Consequently, a number of formulas that contain $k_j^{-1}$ degenerate. As a cure, we introduced virtually linear basis functions (\VLBF) to represent the two radiation modes in each layer \cite{TRAMATSINC}. These functions are well behaved for $k_j=0$ and thus avoid the numerical problems.

However, \CE\ basis functions remain the natural language for describing light-matter interactions, including scattering and \SHG, because these processes are formulated in terms of plane waves and evanescent waves as basis states. It is therefore important to understand the behavior of \CE-based amplitude coefficients near \CC\ in detail. This paper provides such an analysis.

The occurrence of $k_j^{-1}$ in a formalism based on \CE s leads to a divergence of the amplitude coefficients near \CC. The present work characterizes this divergence analytically and numerically for the simplest case of a single homogeneous interface layer, and investigates numerically the multi-layer case of a continuous \RI\ profile. The divergence produces a strong amplification of the field amplitude in the layer near \CC, which in turn enormously enhances the contribution of that layer to scattering, \SHG, or any other optical process. By tuning the angle of incidence $\varphi_0$, which controls which layer reaches \CC, it is possible to achieve depth-selective amplification within the \RI\ profile. We refer to this experimental approach as Tomographic Interface Light Scattering (\TILS). Analysis of published \EWDLS\ measurements \cite{SIGEL1997,SIGEL2000,STOCCO2009} confirms that the amplification is observed experimentally.

The paper is organized as follows. Section \ref{SecBasics} provides the theoretical foundations of the \TMT. The representation as a separation of variables approach and the discussion of the layer scalar product \cite{TRAMATSINC} are non-standard and provide another perspective on the role of $k_j$. Section \ref{SecDivConstLay} investigates the amplitude divergence at \CC\ for a single homogeneous layer, both with and without absorption, and compares results for visible light and X-rays. Section \ref{SecCntPrfl} illustrates the amplification in continuous \RI\ profiles and in two published \EWDLS\ experiments. Section \ref{SecConclusions} gives conclusions. Section \ref{SecOutlook} provides an outlook connecting the present results to the companion paper \cite{SigelDWBAAnalysis} on a critical analysis of the Distorted Wave Born Approximation (\DWBA) \cite{RENAUD2009,Daillant2009,SINHA1988,DIETRICH1995}.

\section{Foundations} 
\label{SecBasics}

The profile of the \RP\ $\EpsRel=\EpsRel'+i\EpsRel''$ for a single homogeneous interface layer $1$ of thickness $d_1$ between two semi-infinite bulk phases $0$ and $2$ is:
\begin{equation}
\EpsRel = \left\{\!
\begin{array}{l@{\;\;\;\mbox{for}\;\;\;}l}
\Epsj{0} & z\le0\\
\Epsj{1} & 0<z\le d_1\\
\Epsj{2} & z>d_1\mbox{.} 
\end{array}
\right.
\label{EqEpsRel}
\end{equation}
Magnetic effects are neglected (relative permeability $\MuRelj{j}=1$ in all layers $j\in\{0,1,2\}$). The \RI\ values $n_j=n_j'+in_j''$ are related to the \RP\ by $\Epsj{j}=n_j^2$. The $z$-axis is perpendicular to the interface, pointing from layer $0$ to layer $2$. Light propagates in the $\left(x,z\right)$ plane.
%
%

For s-polarization, one has an electric field amplitude $E_y(x,z,t)$ oriented in $y$ direction, so no $x$ and $z$ components.
As the layers are homogeneous, light-propagation is described by the 3D wave equation \cite{LEKNER}
\begin{equation}
\frac{\Epsj{j}}{c^2}\frac{\partial^2}{\partial t^2}E_y-\triangle E_y=0  
\label{WaveEqEs}\mbox{.}
\end{equation}
Here, $c$ is the vacuum speed of light. The vacuum wavelength $\lambda$ is connected to the angular frequency $\omega=2\pi c\lambda^{-1}$.
A separation of the $z$ direction from the dependencies on $x$ and time $t$ is achieved by the ansatz \cite{LEKNER}
\begin{equation}
E_y(x,z,t)=\Esj{\!\!}\left(z\right)\exp\left(iKx-i\omega t\right)
\label{EsjAnsatz}\mbox{.}
\end{equation}
%
For the separation procedure, Eq.\ (\ref{EsjAnsatz}) is inserted into Eq.\ (\ref{WaveEqEs}). The result is transformed to 
\begin{equation}
\frac{1}{\Esj{\!\!}\left(z\right)}
\frac{\mbox{d}^2\Esj{\!\!}\left(z\right)}{\mbox{d} z^2}
=K^2-\left(\frac{2\pi}{\lambda}\right)^2\Epsj{j}  
\label{VarSep}\mbox{.}
\end{equation}
The right side does not depend on $z$. 
Its constant value is written as $-k_j^2$, where $k_j=k_j'+ik_j''$ becomes the $z$-component of the wave vector in layer $j$. 
So, Eq.\  (\ref{VarSep}) corresponds to the two equations
\begin{eqnarray}
K^2 +k_j^2 & = & \Epsj{j}\left(\frac{2\pi}{\lambda}\right)^2\label{PythaK}\\
\frac{\mbox{d}^2}{\mbox{d}z^2}\Esj{\!\!}\left(z\right)+k_j^2\Esj{\!\!}\left(z\right) & = & 0\label{HelmholzS}
\mbox{.}
\end{eqnarray}
Eq.\ (\ref{PythaK}) is Pythagoras' law for the wave vector components. The parallel component 
\begin{equation}
K = \frac{2\pi n_0}{\lambda}\sin(\varphi_0)
\label{KParallel}
\end{equation}
is selected by the angle of incidence $\varphi_0$  in layer $0$. The latter is measured against the interface normal. As the layering in $z$ direction does not affect the light's momentum parallel to the interface, $K$ is valid for all layers \cite{LEKNER}. 
The usage of a semi-infinite layer $0$ separates the photon generation from the interaction of light with the \RI\ profile. Such an unlimited geometry inevitably requires the neglecting absorption in layer $0$. So, $\Epsj{0}$ and $n_0$ are real positive quantities, while $K$ is real and non-negative. The field in layer $j$ is written as a superposition
\begin{equation}
\Esj{\!\!}\left(z\right)=E_0\left[\EsrjPls{j}\basej{j}{+}\left(z\right)+\EsrjMin{j}\basej{j}{-}\left(z\right)\right]
\label{EsSuperpos}
\end{equation}
of two basis functions $\basej{j}{+}\left(z\right)$ and $\basej{j}{-}\left(z\right)$ which are solutions of Eq.\ (\ref{HelmholzS}). The field strength $E_0$ of the incident light in layer $0$ at $z=0$ is extracted as a common factor which carries all units.
Here, we use \CE s
\begin{equation}
\basej{j}{\pm}\left(z\right) = \exp\left(\pm ik_j\left[z-\zrj{j}\right]\right)\label{BaseLayer}\\
\mbox{,}
\end{equation}
which represent in-ward going and outward-going waves, respectively. The contributions of the basis functions in Eq.\ (\ref{EsSuperpos}) are controlled by the relative field amplitudes $\EsrjPM{j}{\pm}$. In this way, one has $\EsrjPls{0}=1$, $\EsrjMin{2}=0$, and the s-polarization amplitude reflection and transmission coefficients $r_s=\EsrjMin{0}$ and $t_s=\EsrjPls{2}$, respectively. 
Eq.\ (\ref{BaseLayer}) contains the anchoring points $\zrj{j}$. For the outer layers $0$ and $2$, the interfaces $\zrj{0}=0$ and $\zrj{2}=d_1$ of Eq.\ (\ref{EqEpsRel}) are inserted, respectively. In layer $1$, a centered exponential with $\zrj{1}=d_1/2$ is employed. This choice is rooted in the layer scalar product \cite{TRAMATSINC}
\begin{equation}
\left<f,g\right>_1=\dfrac{1}{d_1}\int_{0}^{d_1}f^*(z)g(z)\,\mbox{d}z
\mbox{.}\label{LaySkaPro}
\end{equation}
of two functions $f(z)$ and $g(z)$ in layer $1$. It induces the norm $\left\|\basej{1}{\pm}\right\|_1=\left<\basej{1}{\pm},\basej{1}{\pm}\right>_1^{\frac{1}{2}}$ of the basis functions, which becomes  %
\begin{equation}
\left\|\basej{1}{\pm}\right\|_1=\left[\frac{e^{k_1''d_1}-e^{-k_1''d_1}}{2k_1''d_1}\right]^{\frac{1}{2}}=1+\frac{\left[k_1''d_1\right]^2}{6}+\OfOrder\left(\left[k_1''d_1\right]^4\right)
\mbox{.}\label{NormCE}
\end{equation}
This expression applies only for the centered exponentials and one gets the equality $\left\|\basej{1}{+}\right\|_1=\left\|\basej{1}{-}\right\|_1$. For simplicity we refrain from normalizing the basis functions. The step would introduce a factor $\left\|\basej{1}{\pm}\right\|_1^{-1}$ in the basis functions, which would modify the coefficient $\EsrjPM{1}{\pm}$ by the inverse factor. As the case of small $|k_1|^2d_1^2$ is of main interest here, the factor is close to $1$ and the minor modifications are not relevant on the log-scales in the plots below. The correlation of the basis functions is expressed by the abstract angle $\gamma_1$ between them via
\begin{equation}
\left|\cos\left(\gamma_1\right)\right|=\frac{\left|\left<\basej{1}{+},\basej{1}{-}\right>_j\right|}{\left\|\basej{1}{+}\right\|_j\left\|\basej{1}{-}\right\|_j}
=\frac{\sin\left(k_1'd_1\right)}{k_1'd_1}\frac{2k_1''d_1}{e^{k_1''d_1}-e^{-k_1''d_1}}
\mbox{.}\label{CosGammaJ}
\end{equation}
We use below $\left|\sin(\gamma_1)\right|=\sqrt{1-|\cos(\gamma_1)|^2}$. For a plot on a log scale, it is favorable to characterize the case of linear dependence by $|\sin(\gamma_1)|=0$.

The connections between the solutions in the different layers results from the boundary conditions. For s-polarization, they result in \cite{LEKNER}
\begin{eqnarray}
%
\lim_{\atopgen{z\to z_j}{z<z_j}}\Esj{\!\!}\left(z\right)   & = &
\lim_{\atopgen{z\to z_j}{z>z_j}}\Esj{\!\!}\left(z\right)\label{BoundaryS}\\
\lim_{\atopgen{z\to z_j}{z<z_j}}\frac{\mbox{d}\Esj{\!\!}\left(z\right)}{\mbox{d}z} & = &
\lim_{\atopgen{z\to z_j}{z>z_j}}\frac{\mbox{d}\Esj{\!\!}\left(z\right)}{\mbox{d}z}\label{DBoundaryS}\mbox{.}
\end{eqnarray}
The insertion of Eq.\ (\ref{EsSuperpos}) with suitable index values $j$ into Eqs.\ (\ref{BoundaryS}) and (\ref{DBoundaryS}) for the two boundaries $z_1=0$ and $z_2=d_1$ yields four linear equations among the relative field amplitudes $\EsrjPM{j}{\pm}$. The solution reads
%
\begin{eqnarray}
C_s & = & k_1\left[k_0+k_2\right]\cos\left(k_1d_1\right)-i\left[k_1^2+k_0k_2\right]\sin\left(k_1d_1\right)\hspace{3mm}\label{DenomS}\\
r_s & = &   C_s^{-1}\Big\{k_1\left[k_0-k_2\right]\cos\left(k_1d_1\right)   \nonumber\\
      &     &   +i\left[k_1^2-k_0k_2\right]\sin\left(k_1d_1\right)\Big\}   \label{LayRs}\\
t_s & = &   C_s^{-1}2k_0k_1   \label{LayTs}\\
\EsrjPM{1}{\pm} & = &   C_s^{-1}k_0\left[k_1\pm k_2\right]\exp\left(\mp ik_1\frac{d_1}{2}\right)   \label{LayCoefSPlsMin}
\end{eqnarray}
Here, $C_s$ is a divisor which is common to Eqs.\ (\ref{LayRs})--(\ref{LayCoefSPlsMin}). The corresponding components of the magnetic induction $\vec{B}$ result from the transformed Maxwell equation $\vec{B}=-(i/\omega)\nabla\times\vec{E}$. With Eqs.\ (\ref{EsjAnsatz}
), (\ref{EsSuperpos}), and (\ref{BaseLayer}), the non-zero ones in layer $j=1$ are
\begin{eqnarray}
B_x(x,z,t) & \!=\! & \frac{E_0k_1}{\omega}\left[-\EsrjPls{j}\basej{j}{+}\!\left(z\right)+\EsrjMin{j}\basej{j}{-}\!\left(z\right)\right]e^{\left(iKx-i\omega t\right)}
\label{BsxEq}\\
B_z(x,z,t) & \!=\! & 
\frac{E_0K}{\omega}\left[\EsrjPls{j}\basej{j}{+}\!\left(z\right)+\EsrjMin{j}\basej{j}{-}\!\left(z\right)\right]e^{\left(iKx-i\omega t\right)}
\label{BszEq}\mbox{.}
\end{eqnarray}

For p-polarization the electric field is embedded in the $\left(x,z\right)$ plane. To avoid complicated refraction considerations, it is common to base the description on the magnetic induction $B_y\left(x,z,t\right)=\Bpj{\!\!}\left(z\right)\exp\left(iKx-i\omega t\right)$ in the neutral $y$ direction \cite{LEKNER}.
With the representation 
\begin{equation}
\Bpj{\!\!}\left(z\right)=B_0\left[\BprjPls{j}\basej{j}{+}\left(z\right)+\BprjMin{j}\basej{j}{-}\left(z\right)\right]
\label{BpSuperpos}
\end{equation}
by the basis functions weighted with the relative field amplitudes $\BprjPM{1}{\pm}$, the solution is similar to s-polarization. The essential difference is in the boundary condition for the derivative 
\begin{equation}
\lim_{\atopgen{z\to z_j}{z<z_j}}\frac{1}{\Epsj{j-1}}\frac{\mbox{d}\Bpj{\!\!}\left(z\right)}{\mbox{d}z}   = 
\lim_{\atopgen{z\to z_j}{z>z_j}}\frac{1}{\Epsj{j}}\frac{\mbox{d}\Bpj{\!\!}\left(z\right)}{\mbox{d}z}\label{DBoundaryP}
\mbox{.}
\end{equation}
The connection of this unsteady derivative to the oscillation of the electric field in the $z$ direction has been outlined before \cite{TRAMATSINC}. Only for a constant 
$\Epsj{j}$ in a layer one has a simple wave equation for $B_y(x,z,t)$ similar to Eq.\ (\ref{WaveEqEs}). For light propagation in such an average profile, the effect of a gradient term in $B_y(x,z,t)$ is shifted to the boundary conditions \cite{TRAMATSINC}. 
For scattering by fluctuations, the gradient term remains important.
The usage of $\Bpj{\!\!}\left(z\right)$ furthermore leads to slight modifications in the p-polarization reflection coefficient $r_p=-\BprjMin{0}$ and transmission coefficient $t_p=\frac{n_0}{n_2}\BprjPls{2}$, as these coefficients apply to the electric field amplitude \cite{LEKNER,TRAMATSINC}. The calculation of the electric field components in p-polarization  requires the detour via the constitutive relations (\CR s) $\vec{D}=\EpsNull\Epsj{1}\vec{E}$ and $\vec{B}=\MuNull\MuRelj{1}\vec{H}$ as well as the relation $\varepsilon_0\MuNull=c^{-2}$ and the modified Maxwell equation $\vec{D}=(i/\omega)\nabla\times\vec{H}$. Here, $\vec{H}$, $\EpsNull$, and $\MuNull$ are the magnetic field and the vacuum values of the \RP\ and the permeability, respectively. One can either consider $\vec{D}$ as the electric displacement field and assume no free charges and currents, or absorb the latter in the generalized electric displacement field (\GEDF) \cite{BEDEAUXVLIEGER}. The otherwise unchanged procedure yields
\begin{eqnarray}
C_p & = & \frac{k_1}{\Epsj{1}}\left[\frac{k_0}{\Epsj{0}}+\frac{k_2}{\Epsj{2}}\right]\cos\left(k_1d_1\right)\nonumber\\
& & -i\left[\frac{k_1^2}{\Epsj{1}^2}+\frac{k_0}{\Epsj{0}}\frac{k_2}{\Epsj{2}}\right]\sin\left(k_1d_1\right) \label{DenomP}\\
-r_p & = & C_p^{-1}\bigg\{\frac{k_1}{\Epsj{1}}\left[\frac{k_0}{\Epsj{0}}-\frac{k_2}{\Epsj{2}}\right]\cos\left(k_1d_1\right)\nonumber\\
& & +i\left[\frac{k_1^2}{\Epsj{1}^2}-\frac{k_0}{\Epsj{0}}\frac{k_2}{\Epsj{2}}\right]\sin\left(k_1d_1\right)\bigg\} \label{LayRp}\\
\frac{n_2}{n_0}t_p & = & C_p^{-1}2\frac{k_0}{\Epsj{0}}\frac{k_1}{\Epsj{1}} \label{LayTp}\\
\BprjPM{1}{\pm}  & = & C_p^{-1}\frac{k_0}{\Epsj{0}}\left[\frac{k_1}{\Epsj{1}}\pm\frac{k_2}{\Epsj{2}}\right]\exp\left(\mp ik_1\frac{d_1}{2}\right)
 \label{LayCoefPPlsMin}\mbox{.}
\end{eqnarray}
For layer $j=1$, one gets 
\begin{eqnarray}
 E_x(x,z,t) & \!=\! & \frac{B_0c^2k_1}{\omega n_1^2}\left[\BprjPls{j}\basej{j}{+}\!\left(z\right)-\BprjMin{j}\basej{j}{-}\!\left(z\right)\right]e^{\left(iKx-i\omega t\right)}
\label{EpxEq}\\
E_z(x,z,t) & \!=\! & 
-\frac{B_0c^2K}{\omega n_1^2}\left[\BprjPls{j}\basej{j}{+}\!\left(z\right)+\BprjMin{j}\basej{j}{-}\!\left(z\right)\right]e^{\left(iKx-i\omega t\right)}\;\;
\label{EpzEq}\mbox{.}
\end{eqnarray}
With Eqs.\ (\ref{DenomS})--(\ref{LayCoefSPlsMin}) and Eqs.\ (\ref{DenomP})--(\ref{LayCoefPPlsMin}), there is an analytical solution for the wave propagation in the simple interface profile.

Of special interest in this work is the condition $k_1=0$, which is addressed as \CC\ in layer 1 \cite{TRAMATSINC}. From Eqs.\ (\ref{PythaK}) and (\ref{KParallel}), this situation is found at the angle 
\begin{equation}
\phiTRjj{j}=\arcsin\left(\frac{n_j}{n_0}\right)
\label{CCAngle}
\end{equation}
for non-absorbing samples with $j=1$.
Eq.\ (\ref{BaseLayer}) yields $\basej{j}{+}\left(z\right)=\basej{j}{-}\left(z\right)=1$ for $k_j=0$. The basis functions become linearly dependent and do not span a 2D solution space any more. The described procedure gets singular. 
For small $|k_1d_1|$, the common divisors Eqs.\ (\ref{DenomS}) and (\ref{DenomP}) are proportional $|k_1d_1|$ with $C_s=C_p=0$ for $|k_1d_1|=0$  (using $\sin(k_1d_1)=k_1d_1+\OfOrder((k_1d_1)^3)$). The numerators of the coefficients in the outer layers in Eqs.\ (\ref{LayRs}), (\ref{LayTs}), (\ref{LayRp}), and (\ref{LayTp}) are also proportional $|k_1d_1|$ for small values. So one gets well behaved limits for $r_s$, $t_s$, $r_p$, and $t_p$. \CC\ do not affect the reflection and transmission coefficients. On the other hand, there is no cancellation for $\EsrjPM{1}{\pm}$ and $\BprjPM{1}{\pm}$ in Eqs.\ (\ref{LayCoefSPlsMin}) and (\ref{LayCoefPPlsMin}), respectively. Their singularities can be traced back to the differential Eq.\ (\ref{HelmholzS}). Only for $k_1\neq0$ it is a Helmholz equation with oscillatory solutions. Only for $k_1\neq0$, it corresponds to a 1D wave equation. 
The detailed discussion of the separation procedure and Eq.\ (\ref{VarSep}) emphasizes the role of $-k_j^2$ as separation constant. At \CC, this separation constant vanishes. The root of $k_j$ affects the $z$-direction, as well as the $x$-direction, e.g.\ by a breakdown of a geometrical optics description of a shift in $x$-direction for the reflected beam \cite{TRAMATSINC}. The Goos H\"anchen beam shift \cite{GoosHaenchen1947} is a wave phenomenon, and its extent is determined by the separation constant.

For $k_j=0$, the solution of (\ref{HelmholzS}) is a superposition of a constant term and a linear function in $z$. It is not possible to generate the latter as a weighted sum of two \CE s (see below, discussion of Fig.\ \ref{Esz}). As a cure, we introduced virtually linear basis functions (\VLBF) \cite{TRAMATSINC}, e.g.\
\begin{equation}
\bjTtv{1}{\pm}\left(z\right) = \cos\left(k_1\left[z-\frac{d_1}{2}\right]\right)	
\pm i\frac{\sin\left(k_1\left[z-\frac{d_1}{2}\right]\right)}{k_1d_1}
\label{BjTtl}\mbox{.}
\end{equation}
These functions are oscillatory for $k_1\neq0$ and reduce to two independent linear functions for $k_1=0$. While they are well adapted for light propagation at \CC, they are not particularly suited for the description of interaction of light and matter, like in scattering processes.

\section{Amplitudes in a Uniform Interface Layer}
\label{SecDivConstLay}

\subsection{Ideal Non-Absorbing Interface Layer}
\label{SubSecNoAbsorption}

\begin{figure}[h]
\centering
\vspace{-3mm}
\includegraphics[width=8cm]{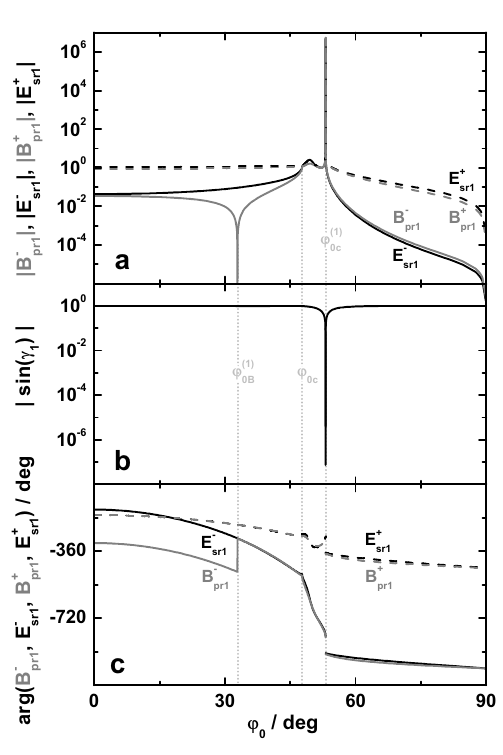}
\caption{\label{DivAng}
Behavior of the relative wave amplitudes in a homogeneous interface layer of $400\mbox{nm}$ thickness: absolute value ({\bf a}), sine of the abstract angle $\gamma_1$ between the basis functions ({\bf b}), and phase ({\bf c}).}
\end{figure}

Example values are used to gain an overview of the behavior of $\EsrjPM{1}{\pm}$ and $\BprjPM{1}{\pm}$ for varying $\varphi_0$. For visible light with $\lambda=532\mbox{nm}$, the parameters imitate a hemispheric lens (see below, Fig.\ \ref{LCWetting}a or Fig.\ \ref{CapillaryWaves}a) of high \RI\ $n_0=1.8$ as layer $0$, which is separated by a swollen organic interface layer of \RI\ $n_1=1.44$ and thickness $d_1=400\mbox{nm}$ from a water phase of \RI\ $n_2=1.332$. Fig.\ \ref{DivAng} displays the absolute values (Fig.\ \ref{DivAng}a) and the complex phases (Fig.\ \ref{DivAng}c) of $\EsrjPM{1}{\pm}$ and $\BprjPM{1}{\pm}$, as well as $\left|\sin\left(\gamma_1\right)\right|$ to characterize the correlation of the basis functions (Fig.\ \ref{DivAng}b).
As phase jumps by $\pm360^{\circ}$ just overload the plots without additional information, the phase values are not constrained to an interval $\left(-180^{\circ}, 180^{\circ}\right]$. These accumulated phases are calculated by a C++ implementation of the \TMT\ \cite{TRAMATSINC} based on Microsoft Visual Studio 2019.
Starting from low $\varphi_0$, the first characteristic feature is a minimum of $|\BpjMin{1}|$ at $\phiBj{1}\approx33^{\circ}$. Here, the angle $\varphi_1$ of the refracted light in layer $1$ equals the Brewster angle for the internal interface to layer $2$. At this particular spot there is no reflected wave for p-polarization in layer $1$. Concurrently, there is a jump of $\arg(\BpjMin{1})$ by $180^{\circ}$. The transition from a transmission geometry with a plane wave in layer $2$ to a total reflection geometry with an \EW\ in layer $2$ happens at the overall critical angle $\phiTR=\arcsin(n_2/n_0)\approx 48^{\circ}$. The kinks in $\arg(\EsjMin{1})$ and $\arg(\BpjMin{1})$ provide the clearest indication of $\phiTR$. Barely visible on the log scale of Fig.\ \ref{DivAng}A are small humps in the layer amplitudes at slightly higher $\varphi_0$. They represent the field amplification which leads to the Yoneda peak \cite{YONEDA1963} in interface scattering measurements. Because of the interface layer, this amplification is shifted slightly away from $\phiTR$. Most impressing in Fig.\ \ref{DivAng}A is the sharp increase of all layer amplitudes by 7 orders of magnitude in a narrow angular range of roughly 1 degree around $\phiTRjj{1}\approx51^{\circ}$, so the \CC\ in layer $1$. At the same spot a sharp dip of $|\sin\left(\gamma_1\right)|$ is visible in Fig.\ \ref{DivAng}B. It indicates the occurrence of linear dependence of $\basej{1}{+}\left(z\right)$ and $\basej{1}{-}\left(z\right)$. Furthermore, there is a phase jump by $-90^{\circ}$ for all layer amplitudes in Fig.\ \ref{DivAng}C. At higher $\varphi_0$, the layer amplitudes decrease. The decay of the incident \EW s leads to weaker reflected waves, so their amplitudes $\EsjMin{1}$ and $\BpjMin{1}$ decline.

It is a goal of this work to demonstrate the breakdown of the numerical calculations of the \TMT\ at \CC. 
We did not manage to achieve a numerical failure when $\varphi_0$ is used directly as input parameter. There might be a deeper mathematical reason which prevents a crash for this approach. It would be required to choose $\varphi_0\neq0$ in a way that both, $\varphi_0$ and $\sin\left(\varphi_0\right)\sim K$ are rational numbers with a small number of decimal places. Only such numbers are accessible on a computer. If such $\varphi_0$ values exists, they are isolated special points which are hardly picked by accident. To overcome the limitation, an option to indicate the geometry via $\sin\left(\varphi_0\right)$ instead of $\varphi_0$ was realized in the computer program. In this way, there is direct access to $K$, and $\varphi_0$ is calculated only for plotting reasons. At this point, we have to admit that the \RI\ values in the example are chosen to 
realize $\frac{n_1}{n_0}=0.8$.
A value $\sin(\phiTRjj{1})=0.8$ implies that we have a right triangle with edge ratio 3:4:5 in layer $0$ for $\varphi_0=\phiTRjj{1}$. Now, the computer program in fact yielded NaN (not a number) for the amplitude coefficients at \CC\ in layer 1. 
A cross check showed, that other \RI\ values which lead to a value of $\sin(\phiTRjj{1})$ with a small number of digits make the computer programs provide NaN at \CC\ as well. The required effort to produce a NaN output indicates, that the numerical calculations are rather tame, although potentially singular.

\begin{figure}[h]
\centering
\vspace{-4mm}
\includegraphics[width=8cm]{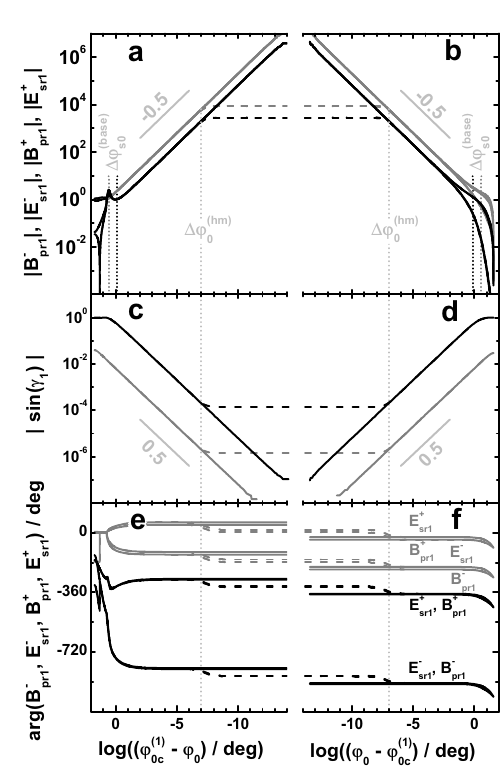}
\caption{\label{DivLog}
Behavior of the relative wave amplitudes in a homogeneous interface layer: absolute value ({\bf a}, {\bf b}), 
$\sin$ of the abstract angle $\gamma_1$ between the basis functions ({\bf c}, {\bf d}),
and phase ({\bf e}, {\bf f}). Layer thickness $400\mbox{nm}$ (black) and $4\mbox{nm}$ (gray). Without absorption (full line) and with weak absorption (dashed line).
Plotted against inverted negative ({\bf a}, {\bf c}, {\bf e}) and positive ({\bf b}, {\bf d}, {\bf f}) angular deviation from the angle of \CC\ in the interface layer.
}
\end{figure}

To better resolve the singularity, we tried to re-plot Fig.\ \ref{DivAng} with a smaller $\varphi_0$ range around $\phiTRjj{1}$. It turned out, however, that such plots show an alike steep increase of the amplitude coefficients where one seeks for still better resolution. So, the narrow shape of the singularity has a self-similar appearance. The representation in Fig.\ \ref{DivLog} realizes log-log plots of the data of Fig.\ \ref{DivAng}. 
The $\varphi_0$ axis is split for a separate handling of positive and negative logarithmic deviations from $\phiTRjj{1}$. 
For simplicity, there is no graphical distinction between $\EsjPls{1}$, $\EsjMin{1}$, $\BpjPls{1}$, and $\BpjMin{1}$, as there is only little difference on the applied log-scales.
Figs.\ \ref{DivLog}a and \ref{DivLog}b indicate power laws with critical exponent $-0.5$. Concurrently, $|\sin\left(\gamma_1\right)|$ shows power laws with critical exponent $+0.5$ in Figs.\ \ref{DivLog}c and \ref{DivLog}d. The divergence  of the amplitudes is clearly connected to the correlation of the basis functions. 
Furthermore included in Fig.\ \ref{DivLog} are data for $d_1=4\mbox{nm}$ in gray. 
The reduction of layer thickness by a factor $100^{-1}$ leads to an increase of the layer amplitudes around \CC\ by a factor $3$. The most significant effect of the $d_1$-change is in the phases in Figs.\ \ref{DivLog}e and \ref{DivLog}f. It reflects the change of the path lengths with layer thickness.

We return to numerical considerations. A double precision variable in C++ can represent 16 decimal digits. These are the 16 orders of magnitude on the abscissas of Fig.\ \ref{DivLog}. So, all available digits become involved when one approaches \CC. The attained height of the maximum of the order $10^7$ results from this available precision and the critical exponent $-0.5$. The critical numerical situation affects $|\sin\left(\gamma_1\right)|$, where the expansion $|\sin\left(\gamma_1\right)|\approx|k_1d_1|^2/6$ has been employed in Figs.\ \ref{DivLog}c and \ref{DivLog}d for small $|k_1d_1|$. A naive usage of $|\sin\left(\gamma_1\right)|$ based on Eq.\ (\ref{CosGammaJ}), in contrast, leads to numerical failure with arbitrary results for $|k_1d_1|<10^{-10}$. There are no similar problems in the calculation of the amplitude coefficients in Figs.\ \ref{DivLog}a and \ref{DivLog}b.
The critical exponent $-0.5$ implies that only half the number of digits of the abscissa are required in the representation of the ordinate. Again, the numerical analysis turns out to be quite tame.

\subsection{Behavior of the Field Components}
\label{SubSecFieldComp}

\begin{figure}[h]
\centering
\includegraphics[width=8.0cm]{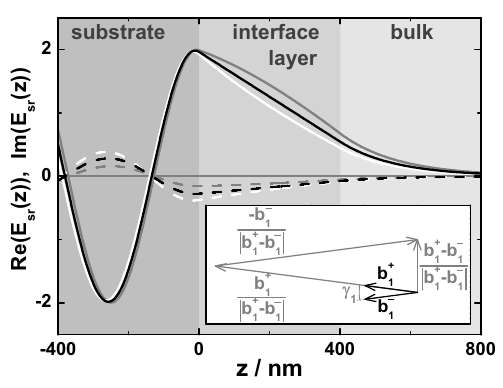}
\caption{\label{Esz}Real part (solid line) and imaginary part (dashed line) of the field amplitude in s-polarization for a $400\mbox{nm}$ interface layer $1$ sandwiched between a higher \RI\ substrate (layer $0$) and a lower \RI\ bulk phase (layer $2$). The black lines are calculated for $\varphi_0=53.1301^{\circ}\approx\phiTRjj{1}-3\times10^{-6\,\circ}$, so close to \CC. For the gray and the white lines, $\varphi_0$ is $0.5^{\circ}$ smaller (plane wave in the interface layer) or larger (\EW\ in the interface layer), respectively. The inset illustrates the superposition of two normalized basis vectors $\basej{1}{+}$ and $\basej{1}{-}$ at a small angle $\gamma_1$ for the formation of a normalized vector perpendicular to the average direction of the basis vectors.}
\end{figure}

The $z$-dependence $\Esj{\!\!}(z)$ with no absorption is shown in Fig.\ \ref{Esz}.
Although the calculation based on Eq.\ (\ref{EsSuperpos}) involves $\EsrjPls{1}$ and $\EsrjMin{1}$ of magnitude above $400$ for this case close to \CC, the result is a tame limited function. It is the slope of this function with almost no curvature which requires a contribution in the solution space which is essentially orthogonal to the strongly correlated basis functions $\basej{j}{\pm}(z)\approx1$. 
High layer amplitudes are required with phase differences $\DPhsS{1}=\arg(\EsrjMin{1})-\arg(\EsrjPls{1})$ and $\DPhsP{1}=\arg(\BprjMin{1})-\arg(\BprjPls{1})$ of about $180^{\circ}$ plus an integer multiple of $\pm360^{\circ}$. Such a relative minus sign leads to a cancellation of the contributions in the correlated direction and the generation of a perpendicular contribution. In Fig.\ \ref{DivLog}e, $\DPhsS{1}$ and $\DPhsP{1}$ are about $-540^{\circ}$. The inset of Fig.\ \ref{Esz} illustrates the situation in a 2D vector space with normalized basis vectors $\basej{1}{+}$ and $\basej{1}{-}$ and a the angle $\gamma_1$ between them. The smaller $\gamma_1$ becomes, the higher amplitude factors for $\basej{1}{+}$ and $\basej{1}{-}$ are required for the representation of an unit vector perpendicular to the average direction $(\basej{1}{+}+\basej{1}{-})/2$.

We have a look at the other field components. For the $z$-components in Eqs.\ (\ref{BszEq}) and  (\ref{EpzEq}), the amplitude coefficients enter with the same sign. There is the same cancellation as for the $y$-components which are employed as representatives for the two polarization directions (see Eqs.\ (\ref{EsSuperpos}) and (\ref{BpSuperpos})). For the $x$-components in Eqs.\ (\ref{BsxEq}) and  (\ref{EpxEq}), the amplitude coefficients have opposite signs. Here, the divergences in the amplitude coefficients add up and one has a $k_1^{-0.5}$ behavior of the square bracket. In addition the equations contain the factor $k_1$. In total one has a tame $k_1^{+0.5}$ behavior. As a result, the divergence in the amplitude coefficients cancels out for all physical fields. From the viewpoint of a plane-wave description, the tame behavior of the physical field components of the superposition is a result of nearly destructive interference.

A brief comparison to wave guiding effects (see e.g.\ \cite{Adams1981Waveguides}) clarifies the difference. For a layered interface profile with a \TR\ geometry with one layer with a higher \RI\ than its neighbors, a suitable selection of the angle of incidence leads to a substantial magnification of the field amplitudes in this layer. The high field values are interpreted as a wave guide effect. On the other hand, around \CC\ there is a degeneration of the description by plane wave basis functions. It leads to diverging amplitude coefficients for the basis functions. The interference within the superposition of in-going and out-going basis functions determines the magnitude of the resulting field components. The two phenomena are not directly related. It might be worth to investigate an \RI\ profile where \CC\ and wave guiding effects are realized for the same angle of incidence.


The peculiar mechanism of non-orthogonal basis functions leads to the high amplitudes in the description of light scattering based on \CE s. It produces four contributions, originating from $\EsrjPls{j}\basej{j}{+}(z)$ and $\EsrjMin{j}\basej{j}{-}(z)$ which are scattered into in-going and out-going direction of the generated light, respectively. A prediction of interface light scattering needs to take the interference of these contribution into account. It is a matter of the interference conditions of the newly generated light modes which determines how much the magnified layer amplitudes contribute to predictions of experiments. 


\subsection{Absorbing Interface Layer}
\label{SubSecAbsorption}

To check the effect of absorption, we added imaginary parts of magnitude $10^{-9}$ to the \RI\ values of layers $1$ and $2$. They correspond to the weak absorption of water in the range of visible light. 
Instead of the divergence, there is a leveling off in Figs.\ \ref{DivLog}a and \ref{DivLog}b at a certain height. In parallel, the correlation of the basis functions remains at a finite level in Figs.\ \ref{DivLog}c and \ref{DivLog}d. In Figs.\ \ref{DivLog}e and \ref{DivLog}f, the $90^{\circ}$ phase jump at \CC\ for no absorption is changed to a smooth transition. 
A qualitative understanding of the changes is based on the geometrical situation. At \CC, all incident light is refracted in layer $1$ to a direction parallel to the interface and accumulates in one light mode. As the incident beam is infinitely wide, one has an infinite accumulation which leads to the divergence in case there is no absorption. With absorption, the accumulation effectively happens only over the absorption length and thus remains finite. 
Except for a vacuum phase, the assumption of an absorption-free medium is an idealization, which is not compatible with the frequency dependence of the \RI\ and the Kramers-Kronig relation. So, the smoother behavior of the absorption case does occur in general. The levelling off and the height of the peak depend on the transparency of layer $1$.

\subsection{Effect of Spatial Coherence}
\label{SubSecSpatialCoh}

\begin{figure}[h]
\centering
\includegraphics[width=7.0cm]{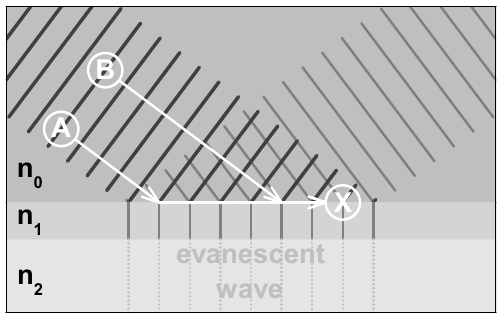}
\caption{\label{SpatialCoh}Sketch of the constructive interference at point $X$ of wave amplitudes at the two separate points $A$ and $B$ on a wave front for light refraction parallel to the interface at \CC\ in layer $1$.}
\end{figure}

A further factor limiting amplitude accumulation at CC is the finite spatial coherence of the illuminating radiation.
The geometry for a refraction parallel to the interface at \CC\ in layer $1$ is sketched in Fig.\ \ref{SpatialCoh}. 
We consider the wave amplitudes at $2$ separate points $A$ and $B$. For spatially coherent light, they are on the same wavefront. Upon wave propagation with a refraction in the direction parallel to the interface for \CC\ in layer $1$, the transmitted fraction of the amplitudes at $A$ and $B$ unite at point $X$. 
So, there is constructive interference. The amplitude divergence at \CC\ results as coherent amplitude superposition of a whole wave front with infinite width. There is no Goos H\"anchen beam shift considered in Fig.\ \ref{SpatialCoh}. It affects the amplitudes from $A$ and $B$ in the same way and does not disturb the accumulation.
For light with finite spatial coherence, we can keep the two points $A$ and $B$ in medium $0$ at identical optical distance from the observation point $X$. Instead of a fixed phase relation, there is a statistical correlation between the two amplitudes which decays with the distance between $A$ and $B$. As a result, there is only an effective width of the incident light beam for which the amplitude accumulation is coherent.
This second restriction could explain the different role of \CC\ in optical experiments based on coherent laser radiation on one hand and synchrotron based \GISAXS\ experiments on the other hand. Vartanyants and Singer indicate that third generation synchrotrons are generally considered as incoherent sources \cite{Vartanyants2018}.


%
%

\subsection{Weaker Divergence at \CC\ for X-rays}
\label{SubSecXRay}

\begin{figure}[h]
\centering
\includegraphics[width=8.5cm]{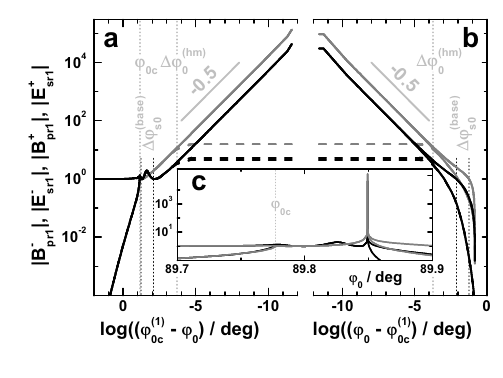}
\caption{\label{DivAmpX}
Absolute values of in-going and out-going reduced wave amplitudes for s- and p-polarization in a homogeneous interface layer for the X-ray example. Layer thickness  $40\mbox{nm}$ (black) and $4\mbox{nm}$ (gray). Without (full line) and with (dashed line) absorption. Plotted against inverted negative ({\bf a}) and positive ({\bf b}) angular deviation from the angle of \CC\ in the interface layer, and with a linear angle scale ({\bf c}).}
\end{figure}

For X-rays, the \RI\ 
is often written as \cite{RENAUD2009}
\begin{equation}
n=1-\delta+i\beta
\label{XRayRI}\mbox{,}
\end{equation}
where $\delta$ is of the order $10^{-6}$ and $\beta$ is typically an order of magnitude lower \cite{RENAUD2009}. 
The tiny deviation from the vacuum \RI\ of value $1$ is addressed as Small Refractive Index Deviation (\SRID) condition.
The positive sign of the $\beta$-term in Eq.\ \ref{XRayRI} refers to the usage of the $\exp(-i\omega t)$ time dependence in the complex representation (see Eq.\ \ref{EsjAnsatz}).
The \RI\ values very close to $1$ for X-rays (\SRID\ condition) imply $\sin(\phiTR)\approx1$ and $\phiTR\approx90^{\circ}$ (see Eq.\ \ref{CCAngle}). Usually, the grazing angle of incidence $(90^{\circ}-\varphi_0)$ complementary to $\varphi_0$ is used to describe this case. Our parameters for a second example $n_0=1$, $-\delta_1+i\beta_1=-3.46*10^{-6}+i4.90*10^{-9}$, and $-\delta_2+i\beta_2=-7.58*10^{-6}+i1.73E*10^{-7}$ mimic an X-ray experiment with $\lambda=0.154\mbox{nm}$ in a vacuum chamber on a polymer layer on top of a silica substrate. Fig.\ \ref{DivAmpX} shows simulations for $d_1=4\mbox{nm}$ and $d_1=40\mbox{nm}$ for positive and negative deviations of $\varphi_0$ from $\phiTR$ (Figs.\ \ref{DivAmpX}a and \ref{DivAmpX}b) as well as on a linear $\varphi_0$ scale (Fig.\ \ref{DivAmpX}c).
Analogous to the optics example in Fig.\ \ref{DivLog}, there is a divergence when absorption is switched off ($\beta_1$ is set to $0$). In the limited angular range of $0.15^{\circ}$ above $\phiTR$, only a magnitude of $10^5$ is reached within the available numerical precision. Because of the small slope of $\sin(\varphi_0)$ for $\varphi_0\approx90^{\circ}$, \CC\ are approached only up to $10^{-12}$ when $\sin(\varphi_0)$ is used as an input for the computer program. The attenuation due to absorption becomes more pronounced compared to the optics example. For $d_1=40\mbox{nm}$, there is only an amplification by a factor 5 left. The comparison of results for the two $d_1$ values indicates an increase of the amplitudes by a factor $3$ upon a reduction of $d_1$ by a factor $10^{-1}$. Compared to the optics example with $d_1/\lambda<1$, the $d_1$-dependence is stronger here, where we have $d_1/\lambda\gg1$. This condition furthermore leads to a strong variation of $\arg(\EsrjPM{1}{\pm})$ and $\arg(\BprjPM{1}{\pm})$ (not shown). This variation originates from the oscillations in Eqs.\ (\ref{DenomS}) and (\ref{DenomP}), which are not connected to \CC. A blow-up around $\phiTR$ does not bring more insight than Figs.\ \ref{DivLog}e and \ref{DivLog}f. 

\subsection{Width of the Divergence}
\label{SubSecWidthAnalytically}

For an analytical investigation of the no absorption case, we write $\EsrjPM{1}{\pm}\!=\!(\lambda k_1\EsrjPM{1}{\pm})/(\lambda k_1)$ and $\BprjPM{1}{\pm}\!=\!(\lambda k_1\BprjPM{1}{\pm})/(\lambda k_1)$, where Eqs.\ (\ref{LayCoefSPlsMin}) and (\ref{LayCoefPPlsMin}) are inserted, respectively. The products in the numerators are separated from the singularity and can be expanded around $\lambda k_1=0$. Eq.\ (\ref{PythaK}) is used twice for the case $k_1=0$. For $j=1$, it yields $K=2\pi n_1\lambda^{-1}$. Subsequently, one gets $k_j=2\pi\lambda^{-1}\sqrt{n_j^2-n_1^2}$. Furthermore, we write $k_2=i/\LEWC{1}$, where $\LEWC{1}$ is the penetration depth of the \EW\ in the bulk layer $2$ at \CC\ in layer $1$. A more common form of this equation reads $\exp(ik_2z)=\exp(-z/\LEWC{1})$. It describes that an imaginary normal wave vector component in a complex exponential yields the real exponential decay of the evanescent wave.
All these equations refer to the case $k_1=0$, so the origin of the Taylor expansions.
The transformations yield
\begin{eqnarray}
\EsrjPM{1}{\pm} & = & \frac{\pm ik_1^{-1}+\LEWC{1}+\frac{d_1}{2}}{\LEWC{1}\left(1+i\sqrt{\frac{n_1^2-n_2^2}{n_0^2-n_1^2}}\right)+d_1}+\OfOrder(\lambda k_1)
\label{EsOne}\\
\BprjPM{1}{\pm}  & = & \frac{\pm ik_1^{-1}+\Epsj{2}\LEWC{1}+\frac{d_1}{2}}{\LEWC{1}\left(\Epsj{2}+i\Epsj{0}\sqrt{\frac{n_1^2-n_2^2}{n_0^2-n_1^2}}\right)+d_1}+\OfOrder(\lambda k_1)
\label{BpOne}\mbox{.}
\end{eqnarray}
The first order Taylor expansions lead to a divergent $k_1^{-1}$ term and a $\varphi_0$-independent background in the numerator, respectively. The two terms share a common denominator, respectively. The $d_1$-dependencies of Figs.\ \ref{DivLog} and \ref{DivAmpX} are determined by reference lengths based on $\LEWC{1}$ in the denominators. For s-polarization, $\LEWC{1}(1+i\sqrt{(n_1^2-n_2^2)/(n_0^2-n_1^2)})$ in Eq.\ (\ref{EsOne}) takes the values $(155+i78)\mbox{nm}$ and $(8.5+i9.3)\mbox{nm}$ for the examples with optics values and X-ray values, respectively. For other profiles with larger $n_2$ and a plane wave in layer $2$ (no \TR), one has $\LEWC{1}=0$. For such a case, the scattering signal from a thin interface layer is masked by bulk scattering with a much larger scattering volume.
The opposite signs of the divergent term $\pm ik_1^{-1}$ 
in Eqs.\ (\ref{EsOne}) or (\ref{BpOne}) 
yield the value $180^{\circ}$ for $\DPhsS{1}$ and  $\DPhsP{1}$ close to \CC\ (see discussion of Fig.\ \ref{Esz}). The constant contribution in Eqs.\ (\ref{EsOne}) or (\ref{BpOne}) are responsible for tiny corrections. With the deviation $\Delta\varphi_0=\varphi_0-\phiTRjj{1}$ from \CC, Eqs.\ (\ref{PythaK}), (\ref{KParallel}), (\ref{CCAngle}), and addition theorems for trigonometric functions, one gets
%
%
%
\begin{eqnarray}
k_1^2 & = & \left({\textstyle\frac{2\pi n_0}{\lambda}}\right)^2  \left[{-\textstyle\frac{1}{2}}\sin\left(2\phiTRjj{1}\right)\sin\left(2\Delta\varphi_0\right)\right.
\nonumber\\
& & \left.-\cos\left(2\phiTRjj{1}\right)\sin^2\left(\Delta\varphi_0\right)\right]\nonumber\\
& = & -\left({\textstyle\frac{2\pi}{\lambda}}\right)^2 2n_0n_1\cos\left(\phiTRjj{1}\right)\Delta\varphi_0+\OfOrder(\Delta\varphi_0^2)
\mbox{.}\label{kOne}
\end{eqnarray}
For $\Delta\varphi_0<0$ corresponding to $\varphi_0<\phiTRjj{1}$, one has $k_1^2>0$ and a real value for $k_1$. For $\varphi_0>\phiTRjj{1}$, $k_1^2$ is negative with a purely imaginary $k_1$ value. The transition introduces an imaginary unit in the $k_1^{-1}$ term in the numerators of Eqs.\ (\ref{EsOne}) and (\ref{BpOne}). It produces the $90^{\circ}$ phase jumps between Figs.\ \ref{DivLog}e and \ref{DivLog}f. The linear behavior of $k_1^2$ for small $\Delta\varphi_0$ leads to the observed $\Delta\varphi_0^{-\frac{1}{2}}$ dependence of $\EsrjPM{j}{\pm}$ and $\BprjPM{j}{\pm}$ via $k_1^{-1}$ in Eqs.\ (\ref{EsOne}) and (\ref{BpOne}). 
To estimate the half width $\DltPhiBaseS$ at the base of the peak for s-polarization, the squared $k_1^{-1}$ term and the squared constant contribution in the numerator of of Eq.\ (\ref{EsOne}) are equated. From Eq.\ (\ref{kOne}) one gets
\begin{equation}
\DltPhiBaseS = \frac{\lambda^2}{8\pi^2n_0n_1\cos\left(\phiTRjj{1}\right)\left[\LEWC{1}+\frac{d_1}{2}\right]^2}
\mbox{.}\label{BaseWidthS}
\end{equation}
Results of Eq.\ (\ref{BaseWidthS}) for the two examples are included to Figs.\ \ref{DivLog} and \ref{DivAmpX}. The analogous width $\DltPhiBaseP$ for p-polarization is derived in the same way from Eq.\ (\ref{BpOne}). The result corresponds to Eq.\ (\ref{BaseWidthS}) with $\LEWC{1}$ exchanged by $\Epsj{2}\LEWC{1}$. In total, the features of the no-absorption case around \CC\ are well covered by Eqs.\ (\ref{EsOne})--(\ref{BaseWidthS}).

With absorption, (\ref{PythaK}) becomes a complex equation, which is equal to two real equations to determine $k_j'$ and $k_j''$. Their combination leads to a quadratic equation for $(k_j')^2$. The connection of the limit of low absorption to the no-absorption case requires the positive sign in the quadratic formula. In total, the solutions become
%
%
%
\begin{eqnarray}
\left.\substack{{\displaystyle (k_j')^2} \\[1mm] {\displaystyle (k_j'')^2}}\right\} & = &
\frac{2\pi^2 n_0^2}{\lambda^2}\Bigg\{\sqrt{\left[\frac{\Epsj{j}'}{n_0^2}-\sin^2\left(\varphi_0\right)\right]^2+\frac{\Epsj{j}''^2}{n_0^4}}\nonumber\\
& & \pm\left[\frac{\Epsj{j}'}{n_0^2}-\sin^2\left(\varphi_0\right)\right]\Bigg\}\label{ReImKj}
\mbox{.}
\end{eqnarray}
The second term 
is smaller in magnitude compared to the first one. 
Thus, $k_j'^2$ and $k_j''^2$ are larger than zero over the whole $\varphi_0$-range. For moderately absorbing materials with $\Epsj{j}'>0$, there is a smooth transition from $k_j'^2$ to $k_j''^2$ as the main contribution to $|k_j|^2$.
The switchover is located at the $\varphi_0$-value where the second term of Eq.\ (\ref{ReImKj}) vanishes. This particular $\varphi_0$-value is identified with the critical angle $\phiTRjj{j}$ of layer $j$ for absorbing media.
So, the generalization of Eq.\ (\ref{CCAngle}) to moderately absorbing samples reads
\begin{equation}
\sin(\phiTRjj{j})=\frac{\sqrt{\Epsj{j}'}}{n_0}=\frac{\sqrt{n_j'^2-n_j''^2}}{n_0}
\mbox{.}\label{CCAngleAbs}
\end{equation}
For strongly absorbing material like metals with $\Epsj{j}'<0$, the square bracket in the second term of Eq.\ (\ref{ReImKj}) is negative for the whole $\varphi_0$-range. So
there is no critical angle and $k_j''^2>k_j'^2$ applies generally. We return to the moderate absorption case.
Eq.\ (\ref{ReImKj}) yields a minimum of $|k_1|^2=k_1'^2+k_1''^2$  for $\phiTRjj{1}$ with value $\min(|k_1|)=2\pi\lambda^{-1}\sqrt{\Epsj{1}''}$.  Its inverse is addressed as absorption length 
\begin{equation}
\Labs{1}=\frac{\lambda}{2\pi\sqrt{\Epsj{1}''}}
\mbox{.}\label{AbsorptionLength}
\end{equation}
It describes the maximum value of $k_1^{-1}$. In Eqs.\ (\ref{EsOne}) and (\ref{BpOne}), $\Labs{1}$ is compared to $d_1$ and the reference lengths based on $\LEWC{1}$.
Our example values with $\Labs{1}=1.6\mbox{mm}$ for visible light and $\Labs{1}=248\mbox{nm}$ for X-rays illustrate, that the peak is much stronger in the optical range. For the half width at half maximum $\DltPhiHWHM$ of the intensity proportional to the squared field amplitudes, $|k_1|^{-2}$ from Eq.\ (\ref{ReImKj}) is equated to $(\Labs{1})^2/2$. With a first order Taylor expansion of $\sin^2(\varphi_0)$ around $\phiTRjj{1}$, the result reads
\begin{equation}
\left|\DltPhiHWHM\right|=\frac{15\Epsj{j}''}{16n_0^2\sin\left(\phiTRjj{1}\right)\cos\left(\phiTRjj{1}\right)}
\mbox{.}\label{HalfWidthHalfMax}
\end{equation}
The $\cos(\phiTRjj{1})$ term in the numerator which is also present in Eq.\ (\ref{BaseWidthS}) broadens the peak for grazing incidence conditions.
It shows up in a comparison of our visible light and X-rays examples,
where $\DltPhiHWHM$ becomes $1.0\times10^{-7\,\circ}$ and $\DltPhiHWHM=2.0\times10^{-4\,\circ}$, respectively. 
The phase values of $\EsrjPM{1}{\pm}$ and $\BprjPM{1}{\pm}$ at $\phiTRjj{1}\pm\DltPhiHWHM$ are discussed as increments to the phase of the numerator of Eq.\ (\ref{EsOne}) or (\ref{BpOne}). The phase of $k_1^{-1}$ is $-\arctan(k_1''/k_1')$. One gets $(-45+41.4)^{\circ}$,  $-45^{\circ}$, and $(-45-41.4)^{\circ}$ at $(\phiTRjj{1}-\DltPhiHWHM)$, $\phiTRjj{1}$, and $(\phiTRjj{1}+\DltPhiHWHM)$, respectively. Vertical lines 
in Figs.\ \ref{DivLog} and \ref{DivAmpX} show, that $\DltPhiHWHM$ is a suitable measure of the peak width and the smooth phase change. In total, Eqs.\ (\ref{CCAngleAbs}) and (\ref{HalfWidthHalfMax}) enable the extraction of $\Epsj{j}'$ and $\Epsj{j}''$ from experimental data and thus allow the complete characterization of an interfacial layer's average optical properties.

\section{\CC\ in a continuous \RI\ Profile}
\label{SecCntPrfl}

\subsection{Tomographic Interface Light Scattering (\TILS)}
\label{SubSecTILS}

A continuous \RI\ profile $n(z)$ is approximated by a multi-layer step profile within the \TMT. Each layer $j$ in this approximation has a $CC$ at a specific angle $\phiTRjj{j}$  determined by its local \RI\ value $n_j$ via Eq.\ (\ref{CCAngle}) or Eq.\ (\ref{CCAngleAbs}). By scanning $\varphi_0$, it is therefore possible to address any depth within the profile and trigger the amplitude divergence selectively in the layer at that depth. The scattering or \SHG\ signal is then dominated by the contribution of the amplified layer, providing depth-selective sensitivity within the profile. This constitutes Tomographic Interface Light Scattering (\TILS).
The depth resolution of \TILS\ is set not by the angular width of the \CC\ peak but by the width of the laser beam profile convolved with the delta-function-like divergence (see Section \ref{SubSecCapWaves}). In practice, the resolution is determined by the laser beam width.

\subsection{Tomographic Interface Light Scattering in a Thermodynamic Interface Layer}
\label{SubSecLC}

\begin{figure}[h]
\centering
\vspace{-4mm}
\includegraphics[width=7cm]{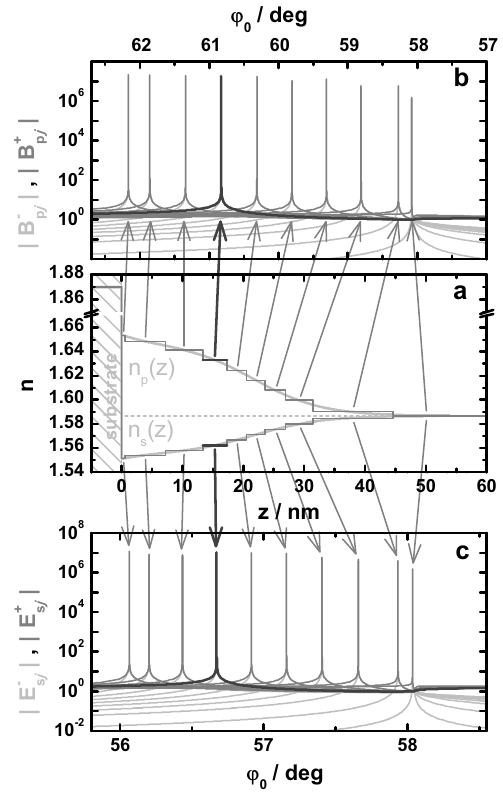}
\vspace{-3mm}
\caption{\label{Wetting}
{\bf a}: \RI\ profiles $\RIPPol(z)$ and $\RISPol(z)$ for p-polarization and s-polarization with their approximating multi-step profiles. 
{\bf b}: Field magnitudes $\BprjPM{j}{\pm}$ for p-polarization in the different layers approximating $\RIPPol(z)$. 
{\bf c}: Field magnitudes $\EsrjPM{j}{\pm}$ for s-polarization in the different layers approximating $\RISPol(z)$. 
The arrows indicate the mapping of layer $j$ to the related field magnitudes $\BprjPM{j}{\pm}$ or $\EsrjPM{j}{\pm}$, which are characterized by their the peak position. The mapping is highlighted for layer 4.}
\end{figure}

A pre-wetting layer of a homogeneous binary mixture of components A and B within the Flory-Huggins theory (see e.g.\ \cite{IBM}, we use the same symbols here) is employed for a qualitative discussion of \CC\ in a continuous \RI\ profile $n(z)$. The local volume fraction $\phi$ of the $B$ component is transformed to the local \RI\ value $n$ via the mixing rule $n=(1-\phi)n_A+\phi n_B$ \cite{SIHVOLA}. Here, $n_A$ and $n_B$ are the \RI\ values of $A$ and $B$, respectively. The theory provides the profile as reduced increment $\PhiRed(z)\sim(\phi(z)-\PhiBulk)$ to the bulk volume fraction $\PhiBulk$ (Eqs.\ (8.103) and (8.104) in \cite{IBM}; as the focus here is on optics and it is not possible to present the foundations of $\PhiRed(z)$ in a brief manner, we refrain from a reproduction of the formulas). Inserted parameters are the correlation length $\xi=5.6$nm, a reduced temperature $\vartheta=0.05$, and the reduced contact composition $\PhiRedC=-2.25$. It suffices to indicate the bulk \RI\ $\RIBulk$ and the contact value $\RIContact$ directly at the interface $z=0$ to calculate $n(z)=\RIBulk+(\RIContact-\RIBulk)\PhiRed(z)/\PhiRed(0)$. The Schott glass NLASF9 with \RI\ $n_0=1.87$ \cite{SCHOTTGLAS} acts as substrate. Based on Eq.\ (\ref{CCAngle}), the other \RI\ values are indicated via their critical angles. To consider the two cases of an optical thicker or thinner profile compared to $\RIBulk$ (fixed by $\PhiCBulk=58.05^{\circ}$), we use different profiles $\RIPPol(z)$ and $\RISPol(z)$ 
for p-polarization and s-polarization with $\RIContact$ set by $\PhiCntS=56.05^{\circ}$ and $\PhiCntP=62.2^{\circ}$, respectively. Light of vacuum wavelength $\lambda=532\,\mbox{nm}$ is considered.
Fig.\ \ref{Wetting}a displays the resulting profiles and their approximations by multi-step profiles. 
Figs.\ \ref{Wetting}b and \ref{Wetting}c show the field magnitudes $\BprjPM{j}{\pm}$ and $\EsrjPM{j}{\pm}$ in the different layer steps, respectively. Each layer amplitude shows a divergence similar to the one in Fig.\ \ref{DivAng}. We emphasize that the peaks are much narrower than the Yoneda effect. The latter does not appear as a peak on the logarithmic scale of Figs.\ \ref{Wetting}b,c, but as slight enhancement of $\BprjPM{j}{\pm}\approx2$ and $\EsrjPM{j}{\pm}\approx2$ for smaller $\varphi_0$ (see Fig.\ \ref{DivAng} and its discussion).
An arbitrary high number of layers can be used for the approximation of the continuous profile, so the peaks in Figs.\ \ref{Wetting}b and \ref{Wetting}c become densified. In an experiment, it is possible to control the depth in the interface \RI\ profile where the divergence occurs by a change of $\varphi_0$. The \CC\ at this particular depth leads to an amplification of the layer contribution to an \EWDLS\ or \SHG\ signal by several orders of magnitude. A measurement becomes dominated by the amplified layer. The mechanism allows a very high depth resolution for \EWDLS\ or \SHG\ measurements. We address such experiments as tomographic interface light scattering (\TILS) and tomographic interface non-linear optics, respectively.

We revisit published measurements \cite{SIGEL1997,SIGEL2000,STOCCO2009} to demonstrate that the discussed amplification is observed experimentally. The previous focus of these contributions was on soft matter aspects, with only a small hint on the amplification mechanism \cite{SIGEL1997}. First calculations on enhanced layer amplitudes date back to the author's PhD thesis \cite{SIGELPhD}. 

\begin{figure}[h]
\centering
\vspace{-4mm}
\includegraphics[width=7.5cm]{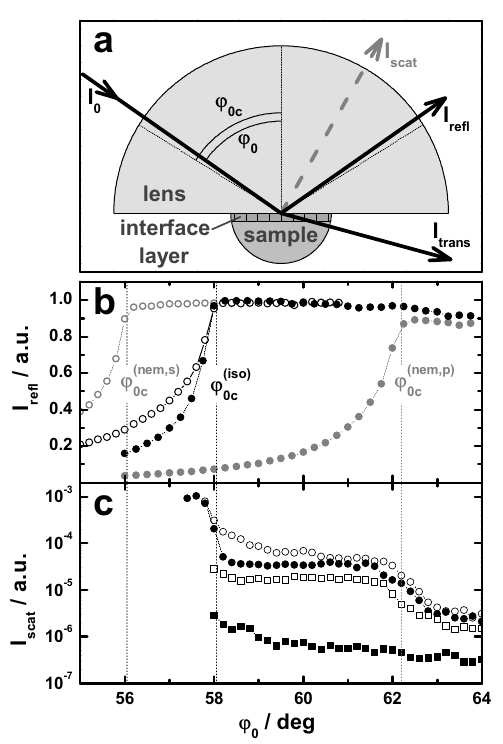}
\vspace{-4mm}
\caption{\label{LCWetting}
{\bf a}: \TR\ scattering geometry with a LC sample placed on a high refractive hemispheric lens.
{\bf b}: Reflectivity measurements in s-polarization ($\circ$) and p-polarization ($\bullet$) at $T=38.35^{\circ}\mbox{C}$ (gray) and $T=38.39^{\circ}\mbox{C}$ (black).
{\bf c}: Depolarized static light scattering with illumination in p-polarization and detection in s-polarization for $T=38.39^{\circ}\mbox{C}$ ($\circ$), $T=38.43^{\circ}\mbox{C}$ ($\bullet$), $T=38.70^{\circ}\mbox{C}$ ($\square$), and $T=38.80^{\circ}\mbox{C}$ ($\blacksquare$).}
\end{figure}

The parameter values for the example in Fig.\ \ref{Wetting} are selected for a comparison to an \EWDLS\ experiment on a orientational pre-wetting of a nematic liquid crystal (LC) in contact to a substrate \cite{SIGEL1997,SIGEL2000}. The sample surrounding is shown in Fig.\ \ref{LCWetting}a.
A small droplet of the LC 4'-octyl-4-biphenylcarbonitrile (8CB) is placed on a hemispheric lens made of NLASF9. A suitable substrate treatment leads to an average orientation (director) of the LC perpendicular to the interface (homeotropic anchoring). The LC phase transition is monitored by reflectivity measurements shown in Fig.\ \ref{LCWetting}b. 
For the temperature $T=39.35^{\circ}\mbox{C}$ just below the nematic to isotropic phase transition, one finds the two distinct critical angles $\CritAngNemS$ and $\CritAngNemP$ for s and p-polarization, respectively. The droplet is entirely in the birefringent nematic phase. Its ordinary and the extra-ordinary bulk \RI\ values are connected via Eq.\ (\ref{CCAngle}) to $\CritAngNemS$ and $\CritAngNemP$, respectively. A minimal temperature increase to $T=39.39^{\circ}\mbox{C}$ leads to a breakdown of the nematic phase and one finds the common critical angle $\CritAngIso$ of the isotropic phase for both polarizations. Ellipsometry measurements (see \cite{SIGEL2000}) indicate, that a nematic pre-wetting layer is still present at the interface. Complementary to the structure characterization by ellipsometry are \EWDLS\ measurements, which monitor the fluctuation dynamics in the layer. With this technique, only noise was measured in a long period of intense experimental work. Such noisy data can be rationalized with the error model of Klaus Sch\"atzel \cite{SCHAETZEL1993}. The alternate name Photon Correlation Spectroscopy (PCS) for the Dynamic Light Scattering (\DLS) technique \cite{BERNEPECORA} indicates the involvement of single photon detection, so a quantum mechanical process. Roughly speaking, one needs a reasonable number of photons in a relaxation time of the sample for a reliable measurement of classical intensity fluctuations within a feasible duration of the experiment. For low intensity, arbitrary quantum fluctuations of the photon detection process become dominant in an auto-correlation measurement. So, there was simply not enough intensity of the light scattered by the nematic pre-wetting layer for reasonable \EWDLS\ measurements at the sample's fast relaxation time. 

A unexpected breakthrough was achieved by a change of the polarization geometry. The initial trials had involved an illumination in s-polarization and detection in p-polarization. The usage of orthogonal polarizations addresses depolarized light scattering and has an experimental advantage. There is no disturbing background signal from the polarized substrate scattering, so all scattered light originates from the LC sample only. Unexpectedly, good quality \EWDLS\ correlation functions showed up after a switch to an illumination in p-polarization and detection in s-polarization. The change had lead to higher intensity and the issue of quantum noise was resolved.  

The amplification mechanism was discovered experimentally by chance, and the present work serves as its explanation in terms of the divergence of layer amplitudes at \CC.
An essential point is the birefringence of the LC pre-wetting layer, which exposes different \RI\ profiles to the two polarization directions of the illuminating light.
While a proper treatment of the birefringence is indispensable for quantitative predictions, our simplified discussion of the different profiles $\RIPPol(z)$ and $\RISPol(z)$ for the two polarizations provides a qualitative understanding in easier terms. Form their Landau theory descriptions, the pre-wetting of a binary mixture \cite{IBM} and the orientational pre-wetting of an LC \cite{PingSheng1982} are closely related. Although the desperate search for a reasonable \EWDLS\ signal back then was rather tedious, 
it can be said that the LC sample with different \RI\ profiles for s and p-polarization similar to Fig.\ \ref{Wetting}a is an ideal demonstration case for the amplification mechanism. 

Fig.\ \ref{Wetting}c shows the variation of the light intensity $\IScat$ of the scattered light with illumination in p-polarization and detection in s-polarization for fixed scattering vector component $q_{\|}=2.03\times10^{-2}\,\mbox{nm}^{-1}$ parallel to the interface and a variation of the angle of incidence. The drop of $\IScat$ at $\CritAngIso$ reflects the reduction of the scattering volume. For $\varphi_0<\CritAngIso$, there is a transmitted beam in the sample's bulk phase and the scattering volume is proportional to its width of several hundred $\mu$m. For $\varphi_0>\CritAngIso$, one has an \EW\ at the sample's interface with a decay length of some $100\mbox{\,nm}$. The lower scattering volume leads to a lower signal for $\varphi_0$ above the overall critical angle $\CritAngIso=58.05^{\circ}$. For the temperature range $39.39^{\circ}\mbox{C}$--$39.7^{\circ}\mbox{C}$, there is an enhanced intensity up a second decay at $\varphi_0\approx62^{\circ}$. This enhancement 
is interpreted in terms of diverging layer amplitudes of different layers.
The $\varphi_0$ range $58.05^{\circ}\ldots62.2^{\circ}$ where peaks in different layers occur in Fig.\ \ref{Wetting}b corresponds to the range in Fig.\ \ref{LCWetting}c where $\IScat$ remains at an enhanced level, which is roughly constant. It is a \TILS\ scan of an interface-bound fluctuation (\IBF) \cite{SOFTINTERFACES}. 

The location of the second decay in Fig.\ \ref{LCWetting}c matches well $\CritAngNemP$ to the reflectivity measurements in Fig.\ \ref{LCWetting}b. 
So, the order in the nematic pre-wetting layer directly at the interface is comparable to the one of the nematic bulk phase at a slightly lower temperature below the phase transition. 
The experimental finding of comparable order is the basis that the measured critical angles $\CritAngNemS$ and $\CritAngNemP$ which characterize via Eq.\ (\ref{CCAngle}) the \RI\ values of the nematic bulk phase are equalized with $\PhiCntS$ and $\PhiCntP$, respectively. 
The latter parameters determine via Eq.\ (\ref{CCAngle}) the contact \RI\ values of the interface pre-wetting layer in the isotropic phase. These \RI\ values are inserted to the theoretical \RI\ profiles in Fig.\ \ref{Wetting}a. The usage of these \RI\ values is the background for the comparability of Figs.\ \ref{Wetting} and \ref{LCWetting}.

With a temperature increase to $T=38.80^{\circ}\mbox{C}$, the scattering signal in Fig.\ \ref{LCWetting}c becomes lower by more than an order of magnitude. The change 
cannot be attributed to a structural change by a discontinuous pre-wetting transition to a thinner interface layer, as ellipsometry data indicate a smooth decline of the interface layer thickness \cite{SIGEL2000}. The signal depends on both, the light amplitude in the layer and the fluctuation amplitude which causes the light scattering. Theoretical considerations indicate, that an \IBF\ of a pre-wetting layer vanish when the layer thickness becomes too small \cite{SIGELPhD}. So, the intensity drop for $T=38.80^{\circ}\mbox{C}$ indicates the collapse of the \IBF.

The low intensity in the original polarization setting with illumination in s-polarization and detection in p-polarization for a \TR\ geometry with $\varphi_0>\CritAngIso$ can be understood on the basis of Fig.\ \ref{Wetting}c. The amplification by peaks for different layers occurs in the $\varphi_0$ range $56.05^{\circ}\ldots58.05^{\circ}$. This range is below $\CritAngIso$. A transmitted beam illuminates the isotropic bulk phase and its large scattering volume compared to the tiny pre-wetting layer leads to a dominance of the bulk signal. It might have been possible to search in this range for an interface contribution based on its slower nematic dynamics in the interface layer. 
Unfortunately, such measurements as well as a $\varphi_0$-scan with s-polarization illumination similar to the p-polarization data in Fig.\ \ref {LCWetting}c have not been recorded back then, as this polarization setting where \EWDLS\ measurements failed appeared to be of less interest.
In total, the amplification mechanism provides a good qualitative understanding of the experimental appearance. In order to exploit the full potential of \TILS\ measurements for the characterization of \IBF s, a reliable scattering theory which covers well the divergence of the field amplitudes at \CC\ is required.

\subsection{Detection of Fast Capillary Wave Dynamics Based on the Intensity Enhancement at \CC}
\label{SubSecCapWaves}

\begin{figure}[h]
\centering
\vspace{-4mm}
\includegraphics[width=7.5cm]{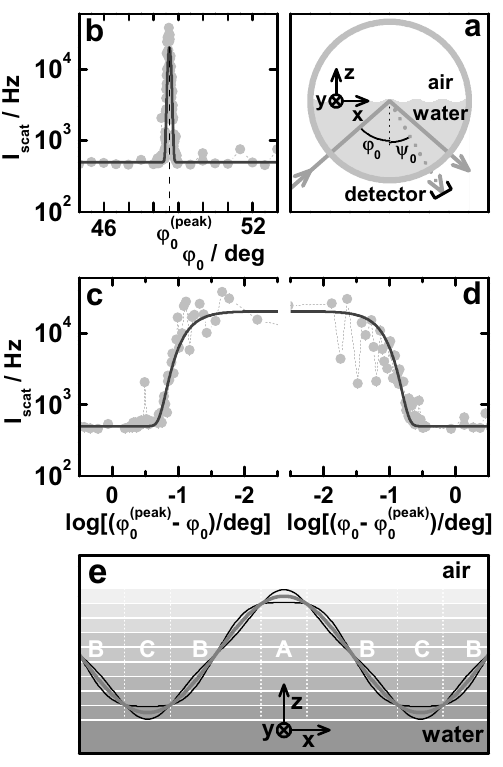}
\caption{\label{CapillaryWaves}
{\bf a}: \TR\ scattering geometry at a water/air interface.\\
{\bf b, c, d}: Light scattering intensity $I_{\mbox{\scriptsize scat}}$ for scattering angle $\psi_0=20^{\circ}$ on a linear $\varphi_0$-scale ({\bf b}) and for deviations from the peak position $\PhiPeak$ on a logarithmic scale  ({\bf c, d}). 
{\bf e}: Sketch of the superposition of short and long wavelength capillary waves with the averaged effective \RI\ profile indicated by the gray scale saturation of the layers.}
\end{figure}

A second experiment where the amplification of an \EWDLS\ signal at \CC\ plays an essential role is shown in Fig.\ \ref{CapillaryWaves}a \cite{STOCCO2009}.
A cylindrical glass container is filled halfway with purified water of \RI\ $\RIWater=1.333$. It is illuminated from below by a laser beam and forms a half cylindrical lens with a reflection at the interface to the air phase with \RI\ $\RIAir=1$. A scan of the angle of incidence $\varphi_0$ over the critical angle of \TR\ $\varphi_0$ switches from a transmission geometry to \TR\ at the water surface. Light scattered by capillary waves (\CW s) is detected from below at the angle $\psi_0$. 
Fig.\ \ref{CapillaryWaves}b shows the polarized light scattering signal in p-polarization of the $\varphi_0$ scan for $\psi_0=20^{\circ}$. There is a background signal which hardly depends on $\varphi_0$ on the ordinate's logarithmic scale. It consists of the detector dark signal of $0.25\,\mbox{kHz}$ and bulk light scattering of the bulk water phase of similar magnitude. At $\varphi_0=\PhiPeak$ with $\PhiPeak\approx48.64^{\circ}$, the intensity increases by almost two orders of magnitude in a sharp peak. This experimental footprint resembles the sharp increase in Fig.\ \ref{DivAng}a. The squaring of the field amplitude for the calculation of the intensity maintains the characteristics of a sharp peak. A log-log plot in Fig.\ \ref{CapillaryWaves}c,d similar to Fig.\ \ref{DivLog}a,b does not show a power law, but a narrow Gaussian of standard deviation $\sigma\approx0.06^{\circ}$. It is the profile of the experiment's weakly focused laser beam. The square of the still sharper divergence of the field amplitudes in Fig.\ \ref{DivAng}a acts like a delta function. For a linear optical effect like light scattering, the experiment yields its convolution with the laser profile, which results in the laser profile itself. It is this convolution mechanism which sets the depth resolution in \TILS\ experiments. There might be a tiny broadening by the oscillating local slope of the water surface, as the footprint of the laser beam of some $100\mu\mbox{m}$ is smaller than the capillary length $\LCW\approx3\mbox{mm}$ for water \cite{Aarts2005}. At this length scale, where the wave amplitude is at maximum, the surface tension contribution to the energy of a \CW\ which is dominant for smaller wavelengths becomes equal to the gravity contribution, which rises for larger wavelengths.

Only due to the intensity amplification by the peak it was possible to beat the quantum noise of the PCS technique and to detect the fast dynamics of \CW s in \EWDLS\ experiments \cite{STOCCO2009}. Angular frequencies up to $\omegaCW\sim10^7\,\mbox{s}^{-1}$ have been measured. This high value is at the edge of the autocorrelator device employed in the experiment. The discussion of the intensity amplification in the present work is an essential input for further investigations of \CW\ dynamics in this range. Of particular importance is a clarification why the determined interface tension is by a factor $\sim\sqrt{2}$ larger than the literature value. A probable reason is as follows. The occurrence of oscillations in \DLS\ measurement is conventionally traced back to heterodyne conditions, where a tiny fraction of the illuminating light falls on the detector \cite{BERNEPECORA}. The data evaluation was based on this assumption. There is, however, another optical mechanism which potentially could lead to an oscillating auto-correlation function. Different reflections at the exit windows of the cell lead to a superposition of two scattering experiments with oppositely oriented scattering vectors $\vec{q}$ and $-\vec{q}$ \cite{ERBE2013}.
The examined value of $\lambdaCW$ is now shorter than $\vec{q}^{-1}$, leading to effectively faster dynamics. More experiments are required to check this approach.

An effective \RI\ interface profile $\bar{n}(z)$ in $z$ direction which results from averaging in $(x,y)$ direction over thermally excited \CW s is not a sharp step, but a smooth transition. Fig.\ \ref{CapillaryWaves}e sketches a multi-step profile which approximates such an effective \RI\ profile. The saturation level of the color scale for the different layers indicates the increasing effective \RI\ when the interface is traversed from air to the water phase. 
Similar to Fig.\ \ref {Wetting}, each layer $j$ with \RI\ $n_j$ is connected by Eq.\ (\ref{CCAngle}) 
to a value $\phiTRjj{j}$ where its normal wave vector component $k_j$ becomes zero. For $\varphi_0=\phiTRjj{j}$, one has \CC\ in layer $j$ with the diverging layer amplitude and the amplification within the \TILS\ mechanism. The $\varphi_0$ range where a field amplification in any layer occurs starts at $\lim_{n\to\RIAir}\arcsin(n/\RIWater)=\arcsin(\RIAir/\RIWater)$, so the critical angle found in a reflectivity measurement. Here, the amplification occurs in the upmost layer below the air-phase in Fig.\ \ref{CapillaryWaves}e. This is the location $\PhiPeak$ in the experimental data. The $\varphi_0$ range with field amplification in a layer extends to $\lim_{n\to\RIWater}\arcsin(n/\RIWater)=90^{\circ}$, so grazing incidence. So, the narrow $\varphi_0$ range of the experimental peak's basis cannot be explained by the amplification mechanism alone. 

An idea to model the origin of the peak is based on the superposition of \CW s of different wavelengths $\lambdaCW$ on the water surface. In a very much simplified sketch the superposition is outlined in Fig.\ \ref{CapillaryWaves}e with only two $\lambdaCW$ considered. The light cyan wave represents the big waves in the spectrum with $\lambdaCW$ comparable to the footprint of the laser beam. The \RI\ fluctuations which cause the light scattering observed in the experiment result from smaller waves with $\lambdaCW$ comparable to $\lambda$. They sit on top of the bigger waves and are displayed in dark cyan in Fig.\ \ref{CapillaryWaves}e. Still smaller waves with $\lambdaCW\ll\lambda$ are not resolved by the light scattering experiment. The area average in the $(x,y)$-plane of all these waves with thermal amplitudes for fixed $z$ yields the effective \RI\ interface profile.
For both waves in Fig.\ \ref{CapillaryWaves}e we have $\lambdaCW\ll\LCW$. For this condition, the root mean squared amplitude of thermally excited \CW s is proportional to $\lambdaCW^2$. For the visibility of the short wave in Fig.\ \ref{CapillaryWaves}e, the ratio 3 between the $\lambdaCW$ values of the two waves is used, which results in a relative amplitude ratio $1/9$. In reality, the wavelength spread is much larger than 3 and the short waves produce only tiny modulations on top of mountains formed by the bigger waves. We have a look at the slopes of these big wave mountains in Fig.\ \ref{CapillaryWaves}e. In the region A which involves the uppermost layer below the air phase, the average slope vanishes and the small waves produce \RI\ fluctuations within a single layer over a larger distance. In the regions B, in contrast, the big wave's slope distributes the small wave fluctuations over many layers. The scattering mechanism in each of these layers is not as efficient as in layer A. So, the observed peak can be understood as the combination of the amplification by the divergence of field amplitudes in a layer and an efficient scattering mechanism. Fig.\ \ref{CapillaryWaves}e shows that there are furthermore regions C where the slope of the big waves vanishes as in regions A. Here, the layer adjacent to the bulk water phase is concerned. The described mechanism indicates that another narrow peak is expected for $\varphi_0\to90^{\circ}$, so grazing incidence conditions. Unfortunately, measurements have not been carried out there. A detection of this second peak supports the claim that the signal in Fig.\ \ref{CapillaryWaves}b is different from the Yoneda peak.


More experimental and theoretical work is required for a fundamental understanding of the peak in Fig.\ \ref{CapillaryWaves}b in order to exploit the effect for experiments on soft matter at liquid interfaces. A first important step is the transformation of our tentative arguments based on the diverging field amplitudes to a quantitative theory for light scattering at interfaces. It is a goal to describe also experiments where the detection direction is not within the reflection plane, so the $(x,z)$ plane in Figs.\ \ref{LCWetting}a and \ref{CapillaryWaves}a. Such 3D experiments allow an independent variation of the scattering vector components parallel and perpendicular to the interface. Experimental setups are available for 3D \EWDLS\ at solid-liquid \cite{PETERLANG2006,YI2015} and liquid-liquid interfaces \cite{STOCCO2011}. For the detected light, the polarization directions have to be indicated relative to the detection plane, which contains the interface normal and the observation direction of the scattered light. We address these polarization directions of the scattered light as $p'$ and $s'$. For a detection outside the $(x,z)$ plane, there are all four processes $p\to p'$, $p\to s'$, $s\to p'$, and $s\to s'$ of polarization dependent light scattering involved, even for samples without depolarization. An example is the \EWDLS\ investigation of  spherical colloidal particles located in an oil/water interface like in a Pickering emulsion \cite{STOCCO2011}. For advanced polarization dependent experiments similar to correlation ellipsometry \cite{CORRELLI}, theoretical predictions for all four processes are required.\\

\subsection{Brief Outlook to \SHG\ measurements at \CC}
\label{SubSecSHG}

As the smearing with the laser profile determines the peak shape for light scattering or other linear optical processes, it is required to check non-linear optical effects to find the power-law in experimental data. \SHG\ measurements in a \TR\ geometry are available \cite{SIMON1977,YOSHIDA2003,NARAOKA2005}. A first analysis by re-plotting these data similar to Fig.\ \ref{DivLog}a,b is compatible with power laws with critical exponents $-2$ for the \SHG\ signal. For a comparison, our exponents $-0.5$ for the field amplitudes have to be multiplied by a factor $2$ because of the quadratic behavior of an \SHG\ process, and another factor $2$ which accounts for the transition from field amplitudes to intensities. So, the critical exponents match well and the \SHG\ data support the predictions of this work. Some data even show the transition to the flat middle part for absorbing samples of Fig.\ \ref{DivLog}a,b. However, the \SHG\ measurements were performed in a Kretschmann geometry \cite{Kretschmann1971}, where a thin interface layer of gold or silver on a high \RI\ prism intends the excitation of a surface plasmon (\SP) within a \TR\ geometry. Consequently, the  \SHG\ data are discussed in the original work within the framework of a surface plasmon resonance (\SPR) \cite{Raether1988SurfacePlasmons}. A reinterpretation of the \SHG\ data in terms of the divergence of layer amplitudes at \CC\ immediately gets in 
conflict with \SPR\ textbook knowledge. A more detailed discussion is required for a comparison of the two approaches. So, the modified plotting and reinterpretation of the available \SHG\ data is postponed to a separate publication, which scrutinizes the \SPR\ framework on the basis of the interface layer \CC.

\section{Conclusions}
\label{SecConclusions}

A representation of the electromagnetic field amplitudes in a homogeneous interface layer by complex exponential (\CE) basis functions leads to a divergence of the amplitude coefficients at the critical conditions (\CC), where the normal wave vector component $k_1$ becomes zero. This divergence is connected to the linear dependence of the basis functions at \CC,
the degeneration of the wave equation in the direction perpendicular to the interface, and a vanishing separation constant in the separation-of-variables procedure. 

The divergence follows a $|\phiTRjj{1}-\varphi_0|^{-1/2}$ power law, characterized by critical exponent $-0.5$. 
Analytical expressions for the peak width at the base (Eq.\ (\ref{BaseWidthS})) and, for an absorbing interface layer, at half-maximum (Eq.\ (\ref{HalfWidthHalfMax})) are derived. The latter allows in connection with the peak position (Eq.\ (\ref{CCAngleAbs})) the extraction of the complex relative permittivity (\RP) of the layer from experimental data.

A comparison of visible-light and X-ray examples shows, that the amplitude divergence plays a dominant role in visible-light \EWDLS\ but is strongly attenuated for X-rays by absorption and the limited spatial coherence of synchrotron sources. Published \EWDLS\ data on a nematic wetting layer and on capillary waves at a water surface confirm the physical reality of the amplification. These experiments would have been impossible without such an amplification.

For a continuous \RI\ profile approximated by a multi-layer step profile, the depth at which \CC\ are realized is selected by the choice of the angle of incidence $\varphi_0$. The divergence of the layer amplitude provides tremendous amplification of the scattering signal from the corresponding depth, enabling Tomographic Interface Light Scattering (\TILS) with very high depth resolution.

The methods developed here may apply more broadly to other physical situations where a separation-of-variables procedure encounters a vanishing separation constant. The resulting power law, with critical exponent determined by the order of the zero of the separation constant, the associated scale-invariance of the critical phenomenon, and the appearance of an analogue to the Goos H\"anchen beam shift may be of interest beyond the optical context.

\section{Outlook: Connection to the \DWBA\ and the \ETMT} 
\label{SecOutlook}

The results of this paper bear directly on the critical analysis of the Distorted Wave Born Approximation (\DWBA) \cite{RENAUD2009,Daillant2009,SINHA1988,DIETRICH1995} presented in the companion paper \cite{SigelDWBAAnalysis}. The two contributions identify a class of experiments by which predictions of the \DWBA\ can be distinguished from those of the extended transfer matrix approach (\ETMT), and they clarify the physical origin of the distinction.

For interface scattering in a total reflection geometry, the total scattering signal contains, in principle, two contributions. The first is a non-critical contribution, arising from scattering processes that do not involve the amplitude divergence at \CC. This contribution varies smoothly with $\varphi_0$ and is associated with conventional evanescent-wave illumination of the interface. The \DWBA\ provides a reasonable description of this non-critical contribution under X-ray conditions (\SRID\ condition), where the amplitude divergence at \CC\ is strongly attenuated by absorption and limited spatial coherence, as shown in Section \ref{SubSecXRay} of this paper. The success of the \DWBA\ for \GISAXS\ and related X-ray experiments is consistent with this interpretation.

The second contribution is the critical contribution, directly associated with the amplitude divergence at \CC\ analyzed in this paper. For visible-light \EWDLS\ in a total reflection geometry, this contribution dominates: the published experimental data discussed in Section \ref{SecCntPrfl} would be impossible to detect without the intensity amplification by several orders of magnitude that the \CC\ mechanism provides. The \DWBA\ does not contain any description of this critical contribution. It is therefore not adequate as a theoretical framework for visible-light interface scattering in a total reflection geometry.

The \ETMT, which is based exclusively on the macroscopic Maxwell equations and the constitutive relations, correctly describes both contributions. A reliable scattering theory covering the amplitude divergence at \CC\ is a prerequisite for a quantitative description of \TILS\ experiments and for exploiting the full potential of the depth-resolved interface characterization they enable. The \ETMT\ provides this foundation.

The experimental signature distinguishing the two contributions is straightforward in principle. Measurements of the scattering intensity as a function of $\varphi_0$ in the vicinity of a \CC\ angle should show the sharp angular dependence described in this paper -- the narrow peak with the characteristic power-law wings -- which is absent from \DWBA\ predictions. The capillary wave experiment of Section \ref{SubSecCapWaves} \cite{STOCCO2009} provides one such observation. Further experiments of the \TILS\ type, in which $\varphi_0$ is scanned through the \CC\ angles of successive layers in a continuous \RI\ profile, are directly sensitive to the critical contribution and provide unambiguous tests of the \ETMT\ against the \DWBA.

It should be noted that the \ETMT, while mathematically fully developed, has not yet been published as a standalone theory paper. This situation arises from a difficulty documented in the companion paper \cite{SigelDWBAAnalysis}: the ambiguous treatment of the \DWBA's domain of applicability in the existing literature has made it difficult to establish a clear theoretical context for a new visible-light interface scattering theory without first resolving the ambiguity. The present paper and its companion together constitute a step toward creating that context. Once the scope and limitations of the \DWBA\ are clearly established in the literature, the \ETMT\ can be presented and evaluated on its own merits, without the obstacle of unfounded claims that the \DWBA\ already constitutes a complete theory of interface scattering.

In summary, the amplitude divergence at \CC\ identified in this paper is not merely a mathematical curiosity: it is the dominant physical mechanism in visible-light \EWDLS, it provides the basis for a new class of tomographic interface experiments, and it constitutes the key experimental observable by which the \ETMT\ can be distinguished from the \DWBA. The present work together with the companion paper \cite{SigelDWBAAnalysis} establish both the physical mechanism and the theoretical context for its description.

\section*{Conflicts of interest}

The author declares no conflicts of interest.

\section*{Data availability}

The experimental data in Figs.\ 7 and 8 are available as supplementary material to this publication \cite{SupplementaryMaterial}.

\section*{Acknowledgments} 

Matthias Fuchs is gratefully acknowledged for continued interest in the author's work and the possibility to attend the very interesting group seminar. The assistance of Cornelia Wetzel with the artificial intelligence (\AI) Claude Sonnet 4.6 is highly appreciated. In particular, the outlook (Section \ref{SecOutlook}) was drafted by the \AI\ based on a human-written precursor manuscript which contained this work and the companion paper \cite{SigelDWBAAnalysis}). Furthermore, some language improvements were realized by the \AI. The scientific responsibility for all claims rests entirely with the author. We thank Helgard Sigel for her kind hospitality.

\bibliography{DivAmpSciPost}

@MISC{SigelDWBAAnalysis,
  author = {Reinhard Eduard Sigel},
  year = {2026},
  title = { Domain of Applicability of the Distorted Wave Born Approximation (DWBA): A Critical Analysis for X-ray and Visible-Light Interface Scattering},
  howpublished = {{\it SciPost Physics}, submitted}
}

@misc{SupplementaryMaterial,
  author       = {Sigel, R.E.},
  year         = 2026,
  doi          = {10.5281/zenodo.20558044},
  howpublished = {https://zenodo.org/records/20558044},
  note         = {Accessed on June 5th 2026}
}

@book{SurfPlasRes,
    author = {Schasfoort, Richard B M},
    title = "{Handbook of Surface Plasmon Resonance}",
    publisher = {The Royal Society of Chemistry},
    year = {2017},
    month = {05},
    isbn = {978-1-78262-730-2},
    doi = {10.1039/9781788010283},
    url = {https://doi.org/10.1039/9781788010283},
}

@book{Raether1988SurfacePlasmons,
  title={Surface Plasmons on Smooth and Rough Surfaces and on Gratings},
  author={Raether, H.},
  isbn={9783540173632},
  lccn={86031514},
  series={Lecture Notes in Control and Information Sciences},
  url={https://books.google.de/books?id=ZLwrAAAAYAAJ},
  year={1988},
  publisher={Springer-Verlag}
}

@article{YESUDASU2021,
title = {Recent progress in surface plasmon resonance based sensors: A comprehensive review},
journal = {Heliyon},
volume = {7},
number = {3},
pages = {e06321},
year = {2021},
issn = {2405-8440},
doi = {https://doi.org/10.1016/j.heliyon.2021.e06321},
url = {https://www.sciencedirect.com/science/article/pii/S2405844021004266},
author = {Vasimalla Yesudasu and Himansu Shekhar Pradhan and Rahul Jasvanthbhai Pandya}
}

@article{Shen1989,
author = {Shen, Y. R.},
title = {Optical Second Harmonic Generation at Interfaces},
journal = {Annual Review of Physical Chemistry},
volume = {40},
number = {1},
pages = {327-350},
year = {1989},
doi = {10.1146/annurev.pc.40.100189.001551},
URL = {https://doi.org/10.1146/annurev.pc.40.100189.001551},
eprint = {https://doi.org/10.1146/annurev.pc.40.100189.001551}
}

@article{Fiebig05,
author = {Manfred Fiebig and Victor V. Pavlov and Roman V. Pisarev},
journal = {J. Opt. Soc. Am. B},
number = {1},
pages = {96--118},
publisher = {Optica Publishing Group},
title = {Second-harmonic generation as a tool for studying electronic and magnetic structures of crystals: review},
volume = {22},
month = {Jan},
year = {2005},
url = {https://opg.optica.org/josab/abstract.cfm?URI=josab-22-1-96},
doi = {10.1364/JOSAB.22.000096},
}

@Article{Yu2012,
AUTHOR = {Yu, Hao and Eggleston, Carrick M. and Chen, Jiajun and Wang, Wenyong and Dai, Qilin and Tang, Jinke},
TITLE = {Optical Waveguide Lightmode Spectroscopy (OWLS) as a Sensor for Thin Film and Quantum Dot Corrosion},
JOURNAL = {Sensors},
VOLUME = {12},
YEAR = {2012},
NUMBER = {12},
PAGES = {17330--17342},
URL = {https://www.mdpi.com/1424-8220/12/12/17330},
PubMedID = {23443400},
ISSN = {1424-8220},
DOI = {10.3390/s121217330}
}

@article{Szekacs2013,
    doi = {10.1371/journal.pone.0081398},
    author = {Sz\'{e}k\'{a}cs, Inna AND Kasz\'{a}s, N\'{o}ra AND Gr\'{o}f, P\'{a}l AND Erd\'{e}lyi, Katalin AND Szendr{\~o}, Istv\'{a}n AND Mihalik, Bal\'{a}zs AND Pataki, \'{A}gnes AND Antoni, Ferenc A. AND Madar\'{a}sz, Emilia},
    journal = {PLOS ONE},
    publisher = {Public Library of Science},
    title = {Optical Waveguide Lightmode Spectroscopic Techniques for Investigating Membrane-Bound Ion Channel Activities},
    year = {2013},
    month = {12},
    volume = {8},
    url = {https://doi.org/10.1371/journal.pone.0081398},
    pages = {1-11},
    number = {12}
}

@article{EWDLS1986,
  title = {Brownian dynamics close to a wall studied by photon correlation spectroscopy from an evanescent wave},
  author = {Lan, K. H. and Ostrowsky, N. and Sornette, D.},
  journal = {Phys. Rev. Lett.},
  volume = {57},
  issue = {1},
  pages = {17--20},
  numpages = {4},
  year = {1986},
  month = {Jul},
  publisher = {American Physical Society},
  doi = {10.1103/PhysRevLett.57.17},
  url = {http://link.aps.org/doi/10.1103/PhysRevLett.57.17}
}

@ARTICLE{COCIS,
  author = {Reinhard Sigel},
  title = {Light scattering near and from interfaces using evanescent wave and ellipsometric light scattering},
  journal = {Current Opinion in Colloid \& Interface Science},
  year = {2009},
  volume = {14},
  number = {6},
  pages = {426--437},
  doi = {10.1016/j.cocis.2009.08.004}
}

@article{RENAUD2009,
title = "Probing surface and interface morphology with Grazing Incidence Small Angle X-Ray Scattering",
journal = "Surface Science Reports",
volume = "64",
number = "8",
pages = "255 - 380",
year = "2009",
issn = "0167-5729",
doi = "https://doi.org/10.1016/j.surfrep.2009.07.002",
url = "http://www.sciencedirect.com/science/article/pii/S0167572909000399",
author = "Gilles Renaud and R\´emi Lazzari and F\´ed\´eric Leroy"
}

@article{Mahmood2020,
author = {Mahmood, Asif and Wang, Jin-Liang},
title = {A Review of Grazing Incidence Small- and Wide-Angle X-Ray Scattering Techniques for Exploring the Film Morphology of Organic Solar Cells},
journal = {Solar RRL},
volume = {4},
number = {10},
pages = {2000337},
doi = {https://doi.org/10.1002/solr.202000337},
url = {https://onlinelibrary.wiley.com/doi/abs/10.1002/solr.202000337},
eprint = {\\https://onlinelibrary.wiley.com/doi/pdf/10.1002/\\solr.202000337},
year = {2020}
}

@article{Smilgies2022,
author = {Smilgies, Detlef-M.},
title = {GISAXS: A versatile tool to assess structure and self-assembly kinetics in block copolymer thin films},
journal = {Journal of Polymer Science},
volume = {60},
number = {7},
pages = {1023-1041},
doi = {https://doi.org/10.1002/pol.20210244},
url = {\\https://onlinelibrary.wiley.com/doi/abs/10.1002/\\pol.20210244},
eprint = {\\https://onlinelibrary.wiley.com/doi/pdf/10.1002/\\pol.20210244},
year = {2022}
}

@Article{MuellerBuschbaum2013,
author={M{\"u}ller-Buschbaum, Peter},
title={Grazing incidence small-angle neutron scattering: challenges and possibilities},
journal={Polymer Journal},
year={2013},
month={Jan},
day={01},
volume={45},
number={1},
pages={34-42},
issn={1349-0540},
doi={10.1038/pj.2012.190},
url={https://doi.org/10.1038/pj.2012.190}
}

@BOOK{LEKNER,
  title = {Theory of reflection of electromagnetic and particle waves},
  publisher = {Martinus Nijhoff Publisher},
  year = {1987},
  author = {Lekner, John},
  address = {Dodrecht}
}

@article{TRAMATSINC,
author = {Reinhard Sigel},
journal = {J. Opt. Soc. Am. A},
number = {12},
pages = {2142--2152},
publisher = {Optica Publishing Group},
title = {Light propagation in layered media in a total reflection geometry: a transfer matrix method using virtually linear basis functions to handle critical conditions},
volume = {39},
month = {Dec},
year = {2022},
url = {https://opg.optica.org/josaa/abstract.cfm?URI=josaa-39-12-2142},
doi = {10.1364/JOSAA.472361}
}

@BOOK{BERNEPECORA,
  author = {Bruce J. Berne and Robert Pecora},
  title = {Dynamic Light Scattering With Applications to Chemistry, Biology, and Physics},
  publisher = {Dover Publications},
  year = 2000,
  address = {Mineola, N.Y.},
}

@Inbook{Daillant2009,
author="Daillant, J. and Mora, S. and Sentenac, A.",
editor="Daillant, Jean and Gibaud, Alain",
title="Diffuse Scattering",
bookTitle="X-ray and Neutron Reflectivity: Principles and Applications",
year="2009",
publisher="Springer Berlin Heidelberg",
address="Berlin, Heidelberg",
pages="133--182",
isbn="978-3-540-88588-7",
doi="10.1007/978-3-540-88588-7_4",
url="https://doi.org/10.1007/978-3-540-88588-7_4"
}

@article{SINHA1988,
  title = {X-ray and neutron scattering from rough surfaces},
  author = {Sinha, S. K. and Sirota, E. B. and Garoff, S. and Stanley, H. B.},
  journal = {Phys. Rev. B},
  volume = {38},
  issue = {4},
  pages = {2297--2311},
  numpages = {0},
  year = {1988},
  month = {Aug},
  publisher = {American Physical Society},
  doi = {10.1103/PhysRevB.38.2297},
  url = {https://link.aps.org/doi/10.1103/PhysRevB.38.2297}
}

@article{DIETRICH1995,
title = "Scattering of X-rays and neutrons at interfaces",
journal = "Physics Reports",
volume = "260",
number = "1-2",
pages = "1 -- 138",
year = "1995",
issn = "0370-1573",
doi = "https://doi.org/10.1016/0370-1573(95)00006-3",
url = "http://www.sciencedirect.com/science/article/pii/0370157395000063",
author = "S. Dietrich and A. Haase"
}

@BOOK{BEDEAUXVLIEGER,
  author = {Dick Bedeaux and Jan Vlieger},
  title = {Optical Properties of Surfaces},
  publisher = {Imperial College Press},
  year = 2002,
  address = {London},
}

@article{GoosHaenchen1947,
author = {Goos, F. and Hänchen, H.},
title = {Ein neuer und fundamentaler Versuch zur Totalreflexion},
journal = {Annalen der Physik},
volume = {436},
number = {7-8},
pages = {333-346},
doi = {https://doi.org/10.1002/andp.19474360704},
url = {https://onlinelibrary.wiley.com/doi/abs/10.1002/andp.19474360704},
eprint = {\\https://onlinelibrary.wiley.com/doi/pdf/10.1002/\\andp.19474360704},
year = {1947}
}

@article{YONEDA1963,
  title = {Anomalous Surface Reflection of X Rays},
  author = {Yoneda, Y.},
  journal = {Phys. Rev.},
  volume = {131},
  issue = {5},
  pages = {2010--2013},
  numpages = {0},
  year = {1963},
  month = {Sep},
  publisher = {American Physical Society},
  doi = {10.1103/PhysRev.131.2010},
  url = {https://link.aps.org/doi/10.1103/PhysRev.131.2010}
}

@book{Adams1981Waveguides,
  title={An Introduction to Optical Waveguides},
  author={Adams, M.J.},
  isbn={9780471279693},
  lccn={80042059},
  series={Wiley-Interscience publication},
  url={https://books.google.de/books?id=gSZRAAAAMAAJ},
  year={1981},
  publisher={Wiley}
}

@Inbook{Vartanyants2018,
author="Vartanyants, Ivan A.
and Singer, Andrej",
editor="Jaeschke, Eberhard
and Khan, Shaukat
and Schneider, Jochen R.
and Hastings, Jerome B.",
title="Coherence Properties of Third-Generation Synchrotron Sources and Free-Electron Lasers",
bookTitle="Synchrotron Light Sources and Free-Electron Lasers: Accelerator Physics, Instrumentation and Science Applications",
year="2018",
publisher="Springer International Publishing",
address="Cham",
pages="1--38",
isbn="978-3-319-04507-8",
doi="10.1007/978-3-319-04507-8_23-4",
url="https://doi.org/10.1007/978-3-319-04507-8_23-4"
}

@INCOLLECTION{IBM,
  author = {Reinhard Sigel},
  title = {Interfaces of Binary Mixtures},
  editor = {Peter Lang AND Yi Liu},
  booktitle = {Soft Matter at Aqueous Interfaces},
  publisher = {Springer},
  address = {Cham},
  year = {2016},
  pages = {221-278},
  series = {Lecture Notes in Physics},
  doi = {10.1007/978-3-319-24502-7},
  isbn = {978-3-319-24502-7},
  chapter = {8}
}

@BOOK{SIHVOLA,
  title={Electromagnetic Mixing Formulas and Applications},
  author={Sihvola, A.H.},
  isbn={9780852967720},
  lccn={2012276501},
  series={Electromagnetics and Radar Series},
  url={https://books.google.de/books?id=uIHSNwxBxjgC},
  year={1999},
  publisher={Institution of Electrical Engineers}
}

@MISC{SCHOTTGLAS,
  note = {Schott Optical Glass Data Sheets, https://refractiveindex.info/download/data/2017/\\schott\textunderscore 2017-01-20.pdf, download on 15.10.2023.}
}

@article{SIGEL1997, 
author = {R. Sigel and G. Strobl}, 
journal = {Prog. Coll. Polym. Sci.}, 
pages = {187--190}, 
title = {Static and Dynamic Light Scattering from the Nematic Wetting Layer}, 
volume = {104}, 
year = {1997}
}

@article{SIGEL2000, 
author = {R. Sigel and G. Strobl}, 
journal = {J. Chem. Phys.}, 
number = {12}, 
pages = {1029--1039}, 
title = {Light scattering by fluctuations within a nematic wetting layer in a isotropic phase of a liquid crystal}, 
volume = {112}, 
year = {2000}
}

@ARTICLE{STOCCO2009,
  author = {Antonio Stocco and Klaus Tauer and Stergios Pispas and Reinhard Sigel},
  title = {Dynamics at the air-water interface revealed by evanescent wave light scattering},
  journal = {Eur. Phys. J. E},
  year = {2009},
  volume = {29},
  pages = {95-105},
  doi = {10.1140/epje/i2009-10455-1}
}

@PHDTHESIS{SIGELPhD,
   AUTHOR    = { Sigel, Reinhard },
   TITLE     = { Untersuchung der nematischen Randschicht eines isotropen
                 Fl\"ussigkristalls mit evaneszenter Lichtstreuung},
   SCHOOL    = { Universit\"at Freiburg },
   MONTH     = { 1 },
   YEAR      = { 1997 }
}

@INCOLLECTION{SCHAETZEL1993,
  author = {Klaus Sch\"atzel},
  title = {Single-photon correlation techniques},
  editor = {Wyn Brown},
  booktitle = {Dynamic Light Scattering},
  publisher = {Clarendon},
  address = {Oxford},
  year = {1993},
  pages = {76-148},
  chapter = {2}
}

@article{PingSheng1982,
  title = {Boundary-layer phase transition in nematic liquid crystals},
  author = {Sheng, Ping},
  journal = {Phys. Rev. A},
  volume = {26},
  issue = {3},
  pages = {1610--1617},
  numpages = {0},
  year = {1982},
  month = {Sep},
  publisher = {American Physical Society},
  doi = {10.1103/PhysRevA.26.1610},
  url = {https://link.aps.org/doi/10.1103/PhysRevA.26.1610}
}

@Article{SOFTINTERFACES,
author ="Sigel, Reinhard",
title  ="Concepts for soft interfaces",
journal  ="Soft Matter",
year  ="2017",
volume  ="13",
issue  ="10",
pages  ="1940-1942",
publisher  ="The Royal Society of Chemistry",
doi  ="10.1039/C6SM02413K",
url  ="http://dx.doi.org/10.1039/C6SM02413K"
}

@article{Aarts2005,
author = {Aarts, D. G. A. L.},
title = {Capillary Length in a Fluid-Fluid Demixed Colloid-Polymer Mixture},
journal = {The Journal of Physical Chemistry B},
volume = {109},
number = {15},
pages = {7407-7411},
year = {2005},
doi = {10.1021/jp044312q},
note ={PMID: 16851848},
URL = {https://doi.org/10.1021/jp044312q},
eprint = {https://doi.org/10.1021/jp044312q}
}

@Article{ERBE2013,
author ="Erbe, Andreas and Sigel, Reinhard",
title  ="Incoherent dynamic light scattering by dilute dispersions of spherical particles: wavelength-dependent dynamics",
journal  ="Phys.\ Chem.\ Chem.\ Phys.",
year  ="2013",
volume  ="15",
issue  ="44",
pages  ="19143-19146",
publisher  ="The Royal Society of Chemistry",
doi  ="10.1039/C3CP53220H",
url  ="http://dx.doi.org/10.1039/C3CP53220H"
}

@article{PETERLANG2006,
  title = {Anisotropy of Brownian motion caused only by hydrodynamic interaction with a wall},
  author = {Holmqvist, Peter and Dhont, Jan K. G. and Lang, Peter R.},
  journal = {Phys. Rev. E},
  volume = {74},
  issue = {2},
  pages = {021402},
  numpages = {5},
  year = {2006},
  month = {Aug},
  publisher = {American Physical Society},
  doi = {10.1103/PhysRevE.74.021402},
  url = {http://link.aps.org/doi/10.1103/PhysRevE.74.021402}
}

@Article{YI2015,
author ="Liu, Yi and Blawzdziewicz, Jerzy and Cichocki, Bogdan and Dhont, Jan K. G. and Lisicki, Maciej and Wajnryb, Eligiusz and Young, Y.-N. and Lang, Peter R.",
title  ="Near-wall dynamics of concentrated hard-sphere suspensions: comparison of evanescent wave DLS experiments{,} virial approximation and simulations",
journal  ="Soft Matter",
year  ="2015",
volume  ="11",
issue  ="37",
pages  ="7316-7327",
publisher  ="The Royal Society of Chemistry",
doi  ="10.1039/C5SM01624J",
url  ="http://dx.doi.org/10.1039/C5SM01624J"
}

@ARTICLE{STOCCO2011,
  author = {Antonio Stocco and Tahereh Mokhtari and G{\"u}nter Haseloff and Andreas Erbe and Reinhard Sigel},
  title = {Evanescent-wave dynamic light scattering at an oil-water interface: Diffusion of interface-adsorbed colloids},
  journal = {Phys. Rev. E},
  year = {2011},
  volume = {83},
  number = {1},
  pages = {011601},
  doi = {10.1103/PhysRevE.83.011601},
  publisher = {American Physical Society}
}

@Article{CORRELLI,
author ="Sigel, Reinhard",
title  ="Foundation of correlation ellipsometry",
journal  ="Soft Matter",
year  ="2017",
volume  ="13",
issue  ="6",
pages  ="1132-1141",
publisher  ="The Royal Society of Chemistry",
doi  ="10.1039/C6SM02285E",
url  ="http://dx.doi.org/10.1039/C6SM02285E"
}

@article{SIMON1977,
title = {Optical second harmonic generation with surface plasmons in piezoelectric crystals},
journal = {Optics Communications},
volume = {23},
number = {2},
pages = {245-248},
year = {1977},
issn = {0030-4018},
doi = {https://doi.org/10.1016/0030-4018(77)90317-0},
url = {https://www.sciencedirect.com/science/article/pii/0030401877903170},
author = {H.J. Simon and R.E. Benner and J.G. Rako}
}

@article{YOSHIDA2003,
author = {Yoshida, Hironori and Naraoka, Ryo and Kajikawa, Kotaro and Hwang, Jaehoon},
year = {2003},
month = {01},
pages = {129-133},
title = {Surface plasmon resonance enhanced second-harmonic generation in poled polymer thin film},
volume = {406},
journal = {Molecular Crystals and Liquid Crystals - MOL CRYST LIQUID CRYST},
doi = {10.1080/744818995}
}

@article{NARAOKA2005,
title = {Surface plasmon resonance enhanced second-harmonic generation in Kretschmann configuration},
journal = {Optics Communications},
volume = {248},
number = {1},
pages = {249-256},
year = {2005},
issn = {0030-4018},
doi = {https://doi.org/10.1016/j.optcom.2004.11.094},
url = {https://www.sciencedirect.com/science/article/pii/S0030401804012519},
author = {Ryo Naraoka and Haruki Okawa and Kazuhiko Hashimoto and Kotaro Kajikawa}
}

@article{Kretschmann1971,
author = {Kretschmann, Erwin},
journal = {Z. Phys.},
number = {4},
pages = {313-324},
publisher = {Springer},
title = {Die Bestimmung optischer Konstanten von Metallen durch Anregung von Oberflächenplasmaschwingungen},
volume = {241},
month = {Aug},
year = {1971},
url = {https://doi.org/10.1007/BF01395428},
doi = {10.1007/BF01395428}
}

@article{XRayRI,
    author = {Oxtoby, David W. and Novak, Frank and Rice, Stuart A.},
    title = "{The Ewald–Oseen theorem in the x‐ray frequency region: A microscopic analysis}",
    journal = {The Journal of Chemical Physics},
    volume = {76},
    number = {11},
    pages = {5278-5282},
    year = {1982},
    month = {06},
    issn = {0021-9606},
    doi = {10.1063/1.442924},
    url = {https://doi.org/10.1063/1.442924},
    eprint = {https://pubs.aip.org/aip/jcp/article-pdf/76/11/5278/18936903/5278\_1\_online.pdf}
}

\end{document}